\documentclass[aps,eqsecnum,a4paper,10pt]{revtex4}
\usepackage{makeidx}
\usepackage{amsmath}
\usepackage{amsfonts}
\usepackage{amssymb}
\usepackage{mathrsfs}
\usepackage[mathscr]{euscript}
\usepackage[dvips]{graphicx}
\usepackage{fancyhdr}

\def\overstrike#1#2{{\setbox0\hbox{$#2$}\hbox to \wd0{\hss
    $#1$\hss}\kern-\wd0\box0}}

\numberwithin{equation}{subsection}

\begin{document}

\title{Wideband pulse propagation: a detailed calculation including Raman processes}
\author{Paul Kinsler}
\affiliation{
  Department of Physics$^*$, Imperial College,
  Prince Consort Road,
  London SW7 2BW, 
  United Kingdom.
}

\begin{abstract}

I present a detailed derivation of 
 wideband optical pulses interacting with 
a Raman transition in the kind of scheme currently used to generate the
ultra broadband light fields needed to create ultrashort pulses.
In contrast to the usual approach using separate field envelopes 
for the pump, Stokes, and anti-Stokes spectral lines, 
I use a {\em single} field envelope. 
This requires the inclusion of few-cycle corrections to the pulse propagation.
The single-field model makes fewer approximations and is mathematically 
(and hence computationally) simpler, 
although it does require greater computational resources to implement.  
The single-field theory reduces to the traditional multi-field 
one using appropriate approximations.

\end{abstract}

\lhead{\includegraphics[height=5mm,angle=0]{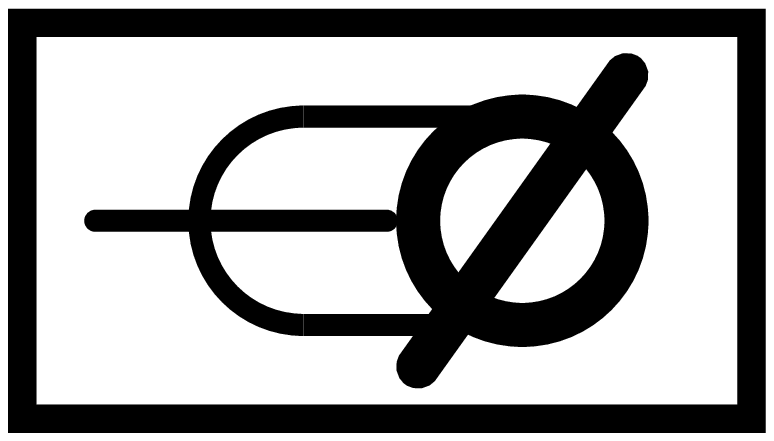}~~WBRAMAN}
\chead{~}
\rhead{
\href{mailto:Dr.Paul.Kinsler@physics.org}{Dr.Paul.Kinsler@physics.org}\\
\href{http://www.kinsler.org/physics/}{http://www.kinsler.org/physics/}
}
%\lfoot{\thesection . \thesubsection; ~~~~ (\yymmdddate\today:\currenttime) }
%\rfoot{{\large {\em Not for redistribution}}}

\date{\today}
\maketitle
\thispagestyle{fancy}

{\large \em
This report should be read along with 
the paper Phys. Rev. {\bf 72}, 033804 (2005) 
``Wideband pulse propagation: single-field and multi-field
approaches to Raman interactions'' by P. Kinsler, G.H.C. New 
\cite{Kinsler-N-2005pra}
for proper context.
This document is primarily intended as a complete (as possible) record 
of the calculational steps that were necessarily 
 abbreviated (or omitted) from that published work.
It is an edited version of a longer document from which on-going work 
 has been excised; 
 and, as a "work in progress", despite my
 best efforts, may contain occasional mistakes.
Please contact me if you have any comments, corrections or queries.
}

\newcommand{\xxref}[1]{{\bf SeeNote:#1:}}
\newcommand{\xxlabel}[1]{[{\bf SeeRef:{#1}}]}

\newcommand{\sech}{{\textrm{ sech}}}

[*] I worked at this institution while doing the bulk of this 
calculation.  My main project was with Jon Marangos \& Prof. GHC New on 
ultrabroadband multifrequency Raman generation, and I was funded 
with money from the EPSRC.

~

\noindent
WWW: QOLS Group 
\href{http://www.qols.ph.ic.ac.uk/}{http://www.qols.ph.ic.ac.uk/} \\
WWW: Physics Dept. 
\href{http://www.ph.ic.ac.uk/}{http://www.ph.ic.ac.uk/} \\
WWW: Imperial College
\href{http://www.ic.ac.uk/}{http://www.ic.ac.uk/} \\
Email: Paul Kinsler 
\href{mailto:Dr.Paul.Kinsler@physics.org}{Dr.Paul.Kinsler@physics.org}\\
Email: G.H.C. New
\href{mailto:g.new@ic.ac.uk}{g.new@ic.ac.uk}

% ----------------------------------------------------------------------
\tableofcontents
%\newpage

\chead{Wideband pulse propagation (Raman)}

% ----------------------------------------------------------------------

\section{Introduction}\label{s-intro}

An important aim of current wideband Raman experiments
is to try to efficiently generate few-cycle pulses
\cite{Harris-S-1998prl,Sokolov-WYYH-2001prl,Hakuta-SKL-2000prl,Sali-MTHM-2004ol}.  
If driven strongly enough, the two-photon Raman transition
modulates the incoming field by adding sidebands separated by the transition
frequency.  Wideband fields are generated as these sidebands generate
sidebands of their own (and so on), thus generating a wide comb of frequency
components separated by the transition frequency. 
If a scheme can be implemented that adjusts the phases of 
each component appropriately, 
then few- or single- cycle optical pulses can be obtained 
(see e.g. \cite{Sokolov-WYYH-2001prl}).  Standard theoretical
treatments of this process split the field into fields components 
centred on the teeth of this comb.  The approach has the advantage that the
components can be modeled reasonably well with slowly varying envelopes, but
of course it has the disadvantage of needing to keep track of a large number
of components.  

In experiments like those of Sali et.al.
\cite{Sali-MTHM-2004ol,Sali-KNMHTM-2005pra}, 
the Raman transition is driven near-resonantly by 
a pair of intense pump pulses about 100fs long; 
compared to the transition frequency of about 130THz, 
the spectra of each pump pulse (and hence the generated sidebands) 
are relatively narrow.  This means that a
multi-component model is still not unreasonable, even if numerical
considerations might demand that the arrays used to store these spectra
overlap in frequency space.  However, if we were to move to shorter pump
pulses, or to a single (much shorter) pump pulse with enough bandwidth to
efficiently excite the transition, we would reach the regime where the
``teeth'' from the spectral comb significantly overlap.  
At this point, not only would we be forced to move from an SVEA 
(Slowly Varying Envelope Approximation) solution of the
field propagation to a more accurate
Generalized Few-cycle Envelope Approximation (GFEA) 
\cite{Kinsler-N-2003pra,Kinsler-FCPP}, 
but the utility of multiple field components becomes questionable.  
In this regime it can be advantageous to treat the field as a single unit,
rather than splitting it into pieces.  
Note that this approach still differs from solutions of Maxwell's
equations such as FDTD (finite difference time domain)\cite{Joseph-T-1997itap}
or PSSD (pseudospectral spatial domain)\cite{Tyyrell-KN-2005jmo}, 
because our single-field is based on a second-order wave equation, 
and uses a convenient choice of carrier function to define a field envelope.

Following these considerations, we now derive a {\em single-field} model for 
Raman generation, and, apart from that notable detail, follow an analogous 
path to that of 
Hickman, Paisner, and Bischel (HPB) \cite{Hickman-PB-1986pra}.
In the model, 
we find that the coupling constants retain an oscillatory behaviour 
at the transition frequency, 
and that it is this that impresses the sideband modulation 
on the propagating field.  
Since the field is not only wideband, but contains significant sideband 
components, we need to propagate this (no longer slowly varying) 
field envelope using the GFEA. 
The necessity of allowing for these can be demonstrated by converting the
single-field model into a multi-field counterpart -- without the
envelope-gradient corrections, we will not get a correct multi-field model.

% ---------------------------------
\subsection{Summary of the theory and the numerical implementation}\label{ss-intro-theorysummary}

We model the  wideband Raman generation process in the following way.  We
specify  the field frequencies ($\omega_i$) of interest, which are usually at
integer spacings of the transition frequency ($\omega_A$) from the main pump
laser frequency ($\omega_0$).  Each of these field components is described
usin a standard envelope theory (i.e. as $A_i(t)$) with a time-history,
allowing us to simulate pulses as they travel through the Raman medium.  The
Raman medium is modelled as a two-level atom using a (its) Bloch vector
$(u,v,w)$, and this Bloch vector is driven by each combination of
spectrally adjacent field components ($\sim \sum A_i A_{-1}$).  Each of the
field components $A_i$ is driven by the atomic polarization ($\sim v$) in
combination with its pair of adjacent field components (i.e. $A_{i-1},
A_{i+1}$).

Each field component evolves as
~
\begin{eqnarray}
    \partial_z A_j
&=&
\frac{\sigma \omega_j \alpha_{12}}
     { \epsilon_0 c_0}
\left\{
  -
  \left[
     v' 
   -
     \imath u'
  \right]
   A_{j+1}
      \exp\left( +\imath (k'_{j+1} -k'_j) z - \imath \Delta t \right)
 +
  \left[
     v' 
   +
     \imath u'
  \right]
    A_{j-1}
      \exp\left( +\imath (k'_{j-1} -k'_j) z + \imath \Delta t \right)
\right\}
\nonumber
\\
&& ~~~~ ~~~~ ~~~~ ~~~~ 
 - 
  \left( k_j - k_0 \right) A_j
\end{eqnarray}

The transition evolves as ($\rho_{12}' \equiv (u',v') $)
\begin{eqnarray}
  \partial_t \rho_{12}' 
&=&
  \left(
     -\gamma_2 + \imath \Delta + 2 \imath g' \sum_j A_j^* A_j 
  \right) 
  \rho_{12}'
+ 
8 \imath f' \sum_j A_{j}  A_{j-1}^*
. w 
. e^{+\imath \left( k_j-k_{j-1}  \right) z - \imath \Delta t } 
\end{eqnarray}

\begin{eqnarray}
\partial_t w
&=&
- \gamma_1 \left( w - w_i \right)  
+  4 \imath f' \sum_j
\left[ 
  A_j^*A_{j+1}
  \rho_{12}' 
         . e^{+\imath \left( k_{j+1} -k_j \right) z - \imath \Delta t } 
-
  A_j^*A_{j-1}
  \rho_{12}'^* 
         . e^{+\imath \left( k_{j-1} -k_j \right) z + \imath \Delta t } 
\right]
\end{eqnarray}

Here $\sigma$ is the number density of the atoms or molecules; 
$\gamma_1$, $\gamma_2$ are the population and polarization decay
rates for the transition; 
$f'$ is the field-transition coupling constant; 
$g'$ is the stark shift coefficient; 
$\omega_j, k_j$ are the frequencies and wavevectors for the field components, 
$\Delta$ gives a rotating frame for $\rho_{12} \rightarrow \rho_{12}'$

Additionally, a Cauchy dispersion is applied to the propagation
of the field components.

Because each of the field components has a time-history, this translates to a
spectral width. In typical cases with roughly nanosecond pulse lengths, the
bandwidth of each component will be tiny compared to the transition frequency,
so there will be large uneventful gaps in the total spectrum.  In contrast, if
the pulse lengths drop to (say) roughly 100 femtoseconds, the bandwidths of
the field components will form a noticeable fraction of the  total spectrum

% ---------------------------------
\subsection{A comment on Cauchy dispersion}\label{ss-intro-comments}

As regards the mismatch terms in the Raman model,
 Geoff New has remarked (email, 20040121) that
 there's a key point about the $\gamma_n$'s that is not made 
 properly in most of the McDonald et al publications; 
 since it is usually said rather enigmatically
 that the $\gamma$'s are "parameterized"  by $\gamma_1$.  
The point is that if one assumes a Cauchy-type law 
 for the refractive index variation, 
 all the gamma's are linked by a recurrence relation, 
 and so you only need to specify one of them, 
 from which all the others will follow.
The point is made properly (to GN's knowledge) only in
 ref \cite{McDonald-NCLL-1998jmo}.

% ----------------------------------------------------------------------

\section{Single-field wideband Raman}\label{s-singlefieldraman}

{\small \it Note: This single field derivation does mean
some of the approximations as to the ``slowness'' of the field 
variation seem somewhat stringent.   However, since a conversion to 
a multi-field model is possible,  it would seem the
field variation constraints are less stringent than they would first 
appear. }

% ---------------------------------
\subsection{Coupled wavefunction equations}\label{ss-single-coupled-wnf}

I start by considering the wave function
$\psi$ of a single molecule (e.g. H$_2$) and the electric field $E$, and
write the time-dependent wave function by expanding it in terms of the
eigenfunctions in the field-free (i.e. $E=0$) case.  
This means I can get the expansion
coefficients by solving for an effective Schr\"oedinger equation that contains
a two-photon Rabi frequency created by means of an interaction term based on a
field-dependent dipole moment.  I assume a dispersionless medium and write all
equations in terms of position $z$ and retarded times $t=t_{lab}-z/c$.  
Here I follow the method of HPB
\cite{Hickman-PB-1986pra}, 
but use only a single $E$ field rather than multiple components.  
Note that HPB use {\em Gaussian} units, so there can appear to be 
inconsistencies when comparing my formulae (in S.I.) to theirs.  

I denote the known molecular eigenfunctions of the 
unperturbed Hamiltonian $H_0$ as $\left| n \right>$, and their
corresponding energies $\hbar W_n$. I want to 
obtain the solution to 
~
\begin{eqnarray}
\left( H_0 + V \right) \psi 
&=& 
  \imath \hbar 
  \frac{\partial \psi}
       {\partial t}
\label{eqn-hamltonian-def}
\\
\textrm{for} ~~~ ~~~
V &=& 
-d E 
\label{eqn-perturbation-def}
\\
\psi &=&
\sum_n
  c_n e^{-\imath W_n t} \left| n \right>
,
\label{eqn-psi-def}
\end{eqnarray}
~
where $d$ is the electronic dipole moment operator and the $c_n$ are a 
set of complex probability amplitudes.

A standard derivation for the equations of motion of the $c_i$
co-efficients proceeds as --
~
\begin{eqnarray}
  \imath \hbar 
  \frac{\partial}
       {\partial t}
  \sum_i c_i e^{-\imath W_i t}
    \left| i \right>
&=&
\left( H_0 + V \right) 
  \sum_j c_j e^{-\imath W_j t}
    \left| j \right>
\\
  \imath \hbar 
  \sum_i 
  \left\{
    -\imath W_i 
    c_i 
    e^{-\imath W_i t}
  +
    e^{-\imath W_i t}
    \frac{\partial c_i}
         {\partial t}
  \right\}
    \left| i \right>
&=&
\left( H_0 + V \right) 
  \sum_j c_j e^{-\imath W_j t}
    \left| j \right>
\\
  \imath \hbar 
\left< n \right|
  \sum_i 
  \left\{
    -\imath W_i 
    c_i 
    e^{-\imath W_i t}
  +
    e^{-\imath W_i t}
    \frac{\partial c_i}
         {\partial t}
  \right\}
    \left| i \right>
&=&
\left< n \right|
\left( H_0 + V \right) 
  \sum_j c_j e^{-\imath W_j t}
    \left| j \right>
\\
  \imath \hbar 
  \left\{
    -\imath W_n 
    c_n 
    e^{-\imath W_n t}
  +
    e^{-\imath W_n t}
    \frac{\partial c_n}
         {\partial t}
  \right\}
&=&
  \hbar
  c_n
  W_n
    e^{-\imath W_n t}
 +
  \sum_j 
    c_j 
    e^{-\imath W_j t}
    \left< n \right|
      -d.E
    \left| j \right>
\\
  \imath \hbar 
  \left\{
    -\imath W_n 
    c_n
  +
    \frac{\partial c_n}
         {\partial t}
  \right\}
&=&
  \hbar
  W_n
  c_n
 -
  \sum_j 
    c_j 
    e^{-\imath W_j t + \imath W_n t}
    \left< n \right|
      d.E
    \left| j \right>
\\
  \imath \hbar 
    \frac{\partial c_n}
         {\partial t}
&=&
 -
  \sum_j 
    c_j 
    e^{-\imath \left( W_j - W_n \right) t}
    \left< n \right|
      d.E
    \left| j \right>
.
\end{eqnarray}

We now use perturbation theory, 
\& $d_{nm} = \left< n \right| \hat{d} \left| n \right>$,
following two independent (but related) strands.

% ---- ---- ----
\subsubsection{CASE (i): Electric field}\label{sss-single-Efield}

This strand leaves the perturbing potential as a function of electric field
$E$, and does not replace it with a carrier-envelope description.  Although
apparently the simplest strategy, it is generally  impractical as it imposes
constraints on the field and other model parameters that are too restrictive
to be useful.

\begin{eqnarray}
\imath \hbar \frac{d c_n}{dt}
&=&
-
\sum_i
  c_i
  d_{ni} E 
  \exp \left[ -\imath \left( W_i - W_n \right) t \right]
\\
\textrm{... assume } c_n, d_{ni} && 
\textrm{vary only slowly, so I can integrate just the exponentials }
\\
c_i
&=&
-
 \frac{1}{ \imath \hbar}
 \sum_j
  c_j
  d_{ij} E 
    \frac{
          \exp \left[ -\imath \left( W_j - W_i \right) t \right]
         }
         {-\imath \left( W_j-W_i \right) }
\\
&=&
\frac{1}{ \hbar}
\sum_j
  c_j
  d_{ij} E
    \frac{
          \exp \left[ -\imath \left( W_j - W_i \right) t \right]
         }
         {W_i-W_j}
\end{eqnarray}

Now substitute the $c_i$ solution into the $d c_n / dt$ equations,
and introduce the shorthand notation $W_{ij} = W_i - W_j$, 
~
\begin{eqnarray}
\imath \hbar \frac{d c_n}{dt}
&=&
-
\sum_i
  \left[
    \frac{1}{ \hbar}
    \sum_j
      c_j
      d_{ij} E
    \frac{
          \exp \left[ -\imath W_{ji} t \right]
         }
         {W_i-W_j}
  \right]
  d_{ni} E 
  \exp \left[ -\imath W_{in} t \right]
\\
&=&
-\frac{1}{ \hbar}
\sum_i
    \sum_j
      c_j
      d_{ij} E
    \frac{
          \exp \left[ -\imath W_{ji} t \right]
         }
         {W_{ij} }
  d_{ni} E 
  \exp \left[ -\imath W_{in} t \right]
\\
&=&
-
      E^2
  \sum_j 
    \frac{1}{\hbar}
    \sum_i
      d_{ni} d_{ij} c_j
    \frac{
          \exp \left[ -\imath W_{jn} t \right]
         }
         {W_i-W_j}
\\
&=&
-
  \sum_j 
    c_j
    \alpha_{nj} E^2
,
\label{eqn-single-E-DcnDt}
\end{eqnarray}
~
where
~
\begin{eqnarray}
\alpha_{nj}
&=&
\frac{1}{\hbar}
\exp \left[ -\imath W_{jn} t \right]
    \sum_i
    \frac{ d_{ni} d_{ij} }
         {W_i-W_j}
\label{eqn-single-E-alpha}
\\
\textrm{UNITS:} ~~~~
\left[ \alpha_{nj} \right]
&\equiv& J^{-1} s^{-1} . Cm . Cm . \left( s^{-1} \right)^{-1}
~~~~ = C^2 m^2 J^{-1}
~~~~ = \left( J V^{-1} \right)^2 m^2 J^{-1}
~~~~ = J m^2 V^{-2}
\\
\textrm{UNITS:} ~~~~
\left[ \frac{\alpha_{nj}}{\hbar} E^2\right]
&\equiv&
J m^2 V^{-2} . J^{-1} s^{-1} . \left( V m^{-1} \right)^2
~~~~ = J . J^{-1}  \times V^{-2} . V^2 \times  m^2 . m^{-2} \times s^{-1}
~~~~ = s^{-1}
\end{eqnarray}

Since we are only interested in the end states $j=1, 2$, between which the
Raman transition occurs, we only need calculate $c_1, c_2$; however we still
retain the sum over all $i$ intermediates states, as they affect  the 
coupling between $1$ and $2$.  The diagonal couplings $\{ \alpha_{nj}, n = j
\}$ are real; but the off-diagonal couplings $\{ \alpha_{nj}, n \neq j \}$
undergo complex oscillations according to the difference in their energy
levels.

Their {\em frequency dependence} is discussed 
after the following subsection; HPB's 
corresponding parameters, which {\em do not oscillate},  were assumed 
to be frequency independent.

% ---- ---- ----
\subsubsection{CASE (ii): Electric Field Envelope}\label{sss-single-Eenvelope}

This strand replaces the electric 
field $E$ with a carrier-envelope 
description, but, unlike HPB, I use only a single carrier-envelope
component rather than a set indexed by some integer $j$.  This 
is necessary, because in the previous strand I ended up 
with coupling constants $\alpha_{nj}$ with strong frequency dependences.

I introduce the envelope and carrier \cite{Gabor-1946jiee} for the field:
~
\begin{eqnarray}
E &=& 
  \left[
    A e^{-\imath \omega_0 t} + A^* e^{+\imath \omega_0 t} 
  \right]
\label{eqn-single-EfromA}
\\
\textrm{so that } ~~~~ ~~~~
\imath \hbar \frac{d c_n}{dt}
&=&
-
\sum_l
  c_l
  d_{nl} 
  \exp \left[ -\imath \left( W_l - W_n \right) t \right]
  \left[
    A e^{-\imath \omega_0 t} + A^* e^{+\imath \omega_0 t} 
  \right]
\\
\textrm{now use~}l\rightarrow i; \textrm{~and assume ~} &c_n, d_{ni}&
\textrm{vary only slowly, so I can integrate just the exponentials }
\\
c_i
&=&
-\frac{1}{ \imath \hbar}
\sum_j
  c_j
  d_{ij} 
  \left[ 
    A
    \frac{
          \exp \left[ -\imath \left( W_j - W_i + \omega_0\right) t \right]
         }
         {-\imath \left( W_j-W_i+\omega_0 \right) }
    +
    A^*
    \frac{
          \exp \left[ -\imath \left( W_j - W_i - \omega_0 \right) t \right]
         }
         {-\imath \left( W_j-W_i-\omega_0 \right) }
  \right]
\\
&=&
\frac{1}{ \hbar}
\sum_j
  c_j
  d_{ij}
  \left[ 
    A
    \frac{
          \exp \left[ -\imath \left( W_j - W_i + \omega_0\right) t \right]
         }
         {W_i-W_j-\omega_0}
    +
    A^*
    \frac{
          \exp \left[ -\imath \left( W_j - W_i - \omega_0 \right) t \right]
         }
         {W_i-W_j+\omega_0}
  \right]
\end{eqnarray}

Note the swap of $\left(W_j-W_i\right)$ to $-\left( W_i-W_j \right)$ 
in the denominators.
Now substitute the $c_i$ solution into the $d c_n / dt$ equations 
(using $l\rightarrow j$),
and introduce the shorthand notation $W_{ij} = W_i - W_j$, 
\begin{eqnarray}
\imath \hbar \frac{d c_n}{dt}
&=&
-
\sum_i
  \left\{
    \frac{1}{ \hbar}
    \sum_j
      c_j
      d_{ij} 
  \left[ 
    A
    \frac{
          \exp \left[ -\imath \left( W_{ji} + \omega_0\right) t \right]
         }
         {W_{ij}-\omega_0}
    +
    A^*
    \frac{
          \exp \left[ -\imath \left( W_{ji} - \omega_0 \right) t \right]
         }
         {W_{ij}+\omega_0}
  \right]
  \right\}
  d_{ni} E 
  \exp \left[ -\imath W_{in} t \right]
\\
&=&
-\frac{1}{ \hbar}
\sum_i
    \sum_j
      c_j
      d_{ij} 
  \left[ 
    A
    \frac{
          \exp \left[ -\imath \left( W_{ji} + \omega_0\right) t \right]
         }
         {W_{ij}-\omega_0}
    +
    A^*
    \frac{
          \exp \left[ -\imath \left( W_{ji} - \omega_0 \right) t \right]
         }
         {W_{ij}+\omega_0}
  \right]
  d_{ni} 
  \left[
    A e^{-\imath \omega_0 t} + A^* e^{+\imath \omega_0 t} 
  \right]
  \exp \left[ -\imath W_{in} t \right]
\\
&=&
-\frac{1}{ \hbar}
\sum_i
    \sum_j
      c_j
      d_{ni} 
      d_{ij} 
  \left[ 
    A
    \frac{
          \exp \left[ -\imath \left( W_{jn} + \omega_0\right) t \right]
         }
         {W_{ij}-\omega_0}
    +
    A^*
    \frac{
          \exp \left[ -\imath \left( W_{jn} - \omega_0 \right) t \right]
         }
         {W_{ij}+\omega_0}
  \right]
  \left[
    A e^{-\imath \omega_0 t} + A^* e^{+\imath \omega_0 t} 
  \right]
\\
&=&
-\frac{1}{ \hbar}
\sum_i
    \sum_j
      c_j
      d_{ni} 
      d_{ij} 
  \left\{
    A^2
    \frac{
          \exp \left[ -\imath \left( W_{jn} + 2 \omega_0\right) t \right]
         }
         {W_{ij}-\omega_0}
   +
    A A^*
    \frac{
          \exp \left[ -\imath W_{jn} t \right]
         }
         {W_{ij}-\omega_0}
\right.
\nonumber
\\
&& ~~~~ ~~~~ ~~~~ ~~~~ ~~~~ ~~~~ ~~~~ ~~~~ 
\left.
   +
    A^* A 
    \frac{
          \exp \left[ -\imath W_{jn} t \right]
         }
         {W_{ij}+\omega_0}
   +
    {A^*}^2
    \frac{
          \exp \left[ -\imath \left( W_{jn} - 2 \omega_0 \right) t \right]
         }
         {W_{ij}+\omega_0}
  \right\}
\\
&\approx&
-\frac{1}{ \hbar}
\sum_i
    \sum_j
      c_j
      d_{ni} 
      d_{ij} 
      A A^*
          \exp \left[ -\imath W_{jn} t \right]
  \left\{
    \frac{1}
         {W_{ij}-\omega_0}
   +
    \frac{1}
         {W_{ij}+\omega_0}
  \right\}
;
~~ \textrm{by discarding the $2 \omega_0$ terms;}
\\
&=&
-\frac{1}{ \hbar}
\sum_i
    \sum_j
      c_j
      d_{ni} 
      d_{ij} 
      A A^*
          \exp \left[ -\imath W_{jn} t \right]
    \frac{W_{ij}+\omega_0 ~~ + ~~ {W_{ij}-\omega_0}}
         { \left( W_{ij}-\omega_0 \right) \left( W_{ij}+\omega_0 \right) }
\\
&=&
-\frac{1}{ \hbar}
\sum_i
    \sum_j
      c_j
      d_{ni} 
      d_{ij} 
      A A^*
          \exp \left[ -\imath W_{jn} t \right]
    \frac{2 W_{ij} }
         {W_{ij}^2-\omega_0^2}
\\
&=&
-    
  A A^*
\sum_j
  c_j
  \sum_i
      \frac{1}{ \hbar}
      d_{ni} 
      d_{ij} 
          \exp \left[ -\imath W_{jn} t \right]
    \frac{2 W_{ij} }
         {W_{ij}^2-\omega_0^2}
\\
&=&
-
\sum_j
  c_j
  \alpha_{nj}'
   ~~ . 2 \left| A \right|^2
\label{eqn-single-A-DcnDt}
\end{eqnarray}
~
where
~
\begin{eqnarray}
\alpha_{nj}'
&=&
      \frac{1}{ \hbar}
          \exp \left[ -\imath W_{jn} t \right]
  \sum_i
      d_{ni} 
      d_{ij} 
    \frac{W_{ij} }
         {W_{ij}^2-\omega_0^2}
\label{eqn-single-A-alpha}
~~~~ ~~~~ = ~~
      \frac{1}{ \hbar}
          \exp \left[ +\imath W_{nj} t \right]
  \sum_i
      d_{ni} 
      d_{ij} 
    \frac{W_{ij} }
         {W_{ij}^2-\omega_0^2}
\end{eqnarray}

These redefined $\alpha_{nj}'$ parameters still oscillate, as they must
because unlike in the HPB derivation, there is no frequency difference between 
field components to cancel with the Raman transition frequency. 
The  coupling also varies with frequency, which is discussed next.

% -------- -------- --------
\subsection{Raman coupling parameters: approximations}

I now discuss two particular (and vital) approximations applied to the 
Raman coupling parameters.

First, note that (as in HPB), I will take the indices $1$ and $2$ to
correspond to the two states involved in the (Raman) transition
of interest; these will be the $0$ and $1$ vibrational (or 
perhaps rotational) levels of the electronic ground state.  Indices
$3$ and above will correspond to (quoting HPB) ``translational 
motion on higher electronic states''.  

{\em Note: I can see that assigning 
these indices to higher electronic states will conveniently 
keep the energy separations for transitions to greater than that of the 
$1 \leftrightarrow 2$  transitions, but it's not so clear to me 
why I can ignore all the higher vibrational (or rotational)
states.}

Since I am interested only in the Raman transition, I specialise the above
equations for the coefficients $c_n$, calculating $c_1$ and $c_2$ only, and
assuming that the $d_{12} = \left< 1 \right| d \left| \right>$ dipole moment is
zero.  This means we will only be including transitions between indices $1$
and $2$ that {\em go via one of the higher states} $j \ge 3$, since we still
allow $d_{1j}, ~d_{2j} \neq 0 ~~ \forall j \ge 3$.  Further, I solve for the
coefficients for the higher states in terms of $c_1$ and $c_2$, in 
an adiabatic approximation justified when $c_1$ and $c_2$ vary only slowly
compared to the exponential terms.

{\em Note: Separate from the oscillations that occur in my coupling
parameters (but not in HPB), there is the issue of {\em frequency dependence}
which applies to both HPB and my parameters.}

For both HPB (their eqn.(15)), and my field-carrier based derivation
(eqn.(\ref{eqn-single-A-alpha})), the presence of the field carrier in the
denominator is helpful.  Since it can reasonably be assumed to be much greater
than the inter(Raman)-level energy differences, the fractional terms will be
correspondingly small, and so the $\alpha_{nj}$ ($\alpha_{nj}'$) parameters
can be assumed to be independent of frequency (or at least nearly so).

This is not in any way obviously (or even partially) true for the no-carrier
$E$ based (eqn.(\ref{eqn-single-E-alpha}), $\alpha_{nj}$) parameters, 
which depend only
on the difference in Raman levels -- at best you might perhaps hope that 
the denominators were approximate multiples of each other.

\subsubsection{Near-zero difference}% $\alpha_{12}'^* - \alpha_{21}'$}

Here I assume that the forward and backward transitions (between 
levels 2 and 1) have nearly the same amplitude.  It is not 
self-evidently true, but HPB must have made an equivalent assumption.
In any case, since $\alpha_{12}'^* - \alpha_{21}'$ is 
the difference of similar terms, it will be smaller by at worst a factor 
of two ($(1+\delta-1) / (1+\delta+1) \sim \delta/2$)
~
\begin{eqnarray}
\alpha_{12}'^* - \alpha_{21}'
&=&
      \frac{1}{ \hbar}
          \exp \left[ +\imath W_{21} t \right]
  \sum_i
      d_{1i}^* 
      d_{i2}^*  
    \frac{ W_{i2} }
         {W_{i2}^2-\omega_0^2}
~~~~
-
      \frac{1}{ \hbar}
          \exp \left[ -\imath W_{12} t \right]
  \sum_i
      d_{2i} 
      d_{i1} 
    \frac{ W_{i1} }
         {W_{i1}^2-\omega_0^2}
\\
&=&
      \frac{1}{ \hbar}
  \sum_i
      d_{1i}^*  
      d_{i2}^*  
    \left[
      \frac{ W_{i2} }
           {W_{i2}^2-\omega_0^2} 
     -
      \frac{ W_{i1} }
           {W_{i1}^2-\omega_0^2}
    \right]
          \exp \left[ +\imath W_{21} t \right]
\\
&\approx&
0
\end{eqnarray}

This approximation allows me to 
equate $\alpha_{21}'$ to $\alpha_{12}'^*$, and hence
change $- \alpha_{21}' c_1^* c_1
+ \alpha_{12}'^* c_2 c_2^* \longrightarrow 
\alpha_{21}' \left( c_2^* c_2
- c_1^* c_1 \right)$ in the $\rho_{12}$ equation below.  This is a vital step
in getting to a simple form equivalent to the Bloch equations.

\subsubsection{Sum is double}% $\alpha_{12}' + \alpha_{21}'^*$}

This follows from the above assumption (as per HPB) that the forward and 
backward transitions (between levels 2 and 1) have nearly the same amplitude.  
This approximation allows me to replace $\alpha_{21}'$ 
with  $\alpha_{12}'$ 
in the $w$ equation below, which simplifies my notation
and makes the coupling term match that in the $\rho_{12}$ equation, 
important in getting to true Bloch equations.

~
\begin{eqnarray}
\alpha_{12}' + \alpha_{21}'^*
&=&
      \frac{1}{ \hbar}
          \exp \left[ -\imath W_{21} t \right]
  \sum_i
      d_{1i} 
      d_{i2} 
    \frac{ W_{i2} }
         {W_{i2}^2-\omega_0^2}
~~~~
+
      \frac{1}{ \hbar}
          \exp \left[ +\imath W_{12} t \right]
  \sum_i
      d_{2i}^* 
      d_{i1}^* 
    \frac{ W_{i1} }
         {W_{i2}^2-\omega_0^2}
\\
&=&
      \frac{1}{ \hbar}
  \sum_i
      d_{1i} 
      d_{i2} 
    \left[
      \frac{ W_{i2} }
           {W_{i2}^2-\omega_0^2} 
     +
      \frac{ W_{i1} }
           {W_{i2}^2-\omega_0^2}
    \right]
          \exp \left[ -\imath W_{21} t \right]
\\
&\approx&
\bar{\alpha}_{12}  
  e^{-\imath W_{21} t}
\label{eqn-alpha-bar}
\end{eqnarray}

\subsubsection{Sum is cosine}% $\alpha_{12}' + \alpha_{21}'$ }

I do not reduce the exponential sum to
a (trig) cosine function, for potential use  later when matching and/or
differencing exponentials. However, this sum is does not occur, 
so the $\bar{a}_{12}$ is never used.
~
\begin{eqnarray}
\alpha_{12}' + \alpha_{21}'
&=&
      \frac{1}{ \hbar}
          \exp \left[ -\imath W_{21} t \right]
  \sum_i
      d_{1i} 
      d_{i2} 
    \frac{ W_{i2} }
         {W_{i2}^2-\omega_0^2}
~~~~
+
      \frac{1}{ \hbar}
          \exp \left[ -\imath W_{12} t \right]
  \sum_i
      d_{2i} 
      d_{i1} 
    \frac{ W_{i1} }
         {W_{i2}^2-\omega_0^2}
\\
&=&
      \frac{1}{ \hbar}
  \sum_i
      d_{1i} 
      d_{i2} 
    \left[
      \frac{ W_{i2} }
           {W_{i2}^2-\omega_0^2} 
          \exp \left[ -\imath W_{21} t \right]
     +
      \frac{ W_{i1} }
           {W_{i2}^2-\omega_0^2}
          \exp \left[ +\imath W_{21} t \right]
    \right]
\\
&\approx&
\bar{a}_{12}  \left[ e^{-\imath W_{21} t} + e^{+\imath W_{21} t} \right]
\label{eqn-alpha-bar-B}
\end{eqnarray}

\subsubsection{Stark Shifts}%: $\alpha_{11}' - \alpha_{22}'^*$}

Here I calculate the Stark shift term
~
\begin{eqnarray}
\alpha_{11}' - \alpha_{22}'^*
&=&
      \frac{1}{ \hbar}
          \exp \left[ -\imath W_{11} t \right]
  \sum_i
      d_{1i} 
      d_{i1} 
    \frac{ W_{i1} }
         {W_{i1}^2-\omega_0^2}
~~~~
-
      \frac{1}{ \hbar}
          \exp \left[ +\imath W_{22} t \right]
  \sum_i
      d_{2i}^* 
      d_{i2}^* 
    \frac{ W_{i2} }
         {W_{i2}^2-\omega_0^2}
\\
&=&
      \frac{1}{ \hbar}
  \sum_i
  \left[
      d_{i1}^*d_{i1} 
    \frac{ W_{i1} }
         {W_{i1}^2-\omega_0^2}
 - 
      d_{2i}^* d_{2i}
    \frac{ W_{i2} }
         {W_{i2}^2-\omega_0^2}
  \right]
\label{eqn-alpha-stark}
\end{eqnarray}

Simplify assuming $W_{21} \ll W_{i1}, W_{i2}$, while
assuming $W_{i1}/\left(W_{i1}^2-\omega_0^2 \right) \sim 1$; or, more
accurately, that 
~
\begin{eqnarray}
1 
&\gg&
      \frac{W_{21}^n}{ W_{i1}^n}
    \frac{ W_{i1}^2 }
         {W_{i1}^2-\omega_0^2}
~~~~ ~~~~ \textrm{for } ~~~~ n ~ = ~~1, ~2
,
\end{eqnarray}

and keeping only the terms 
first order in $W_{21}/W_{i1}, W_{21}/W_{i2}$, etc
~
\begin{eqnarray}
\alpha_{11}' - \alpha_{22}'^*
&=&
      \frac{1}{ \hbar}
  \sum_i
  \left\{
      \left| d_{1i} \right|^2
    \frac{ W_{i1} }
         {W_{i1}^2-\omega_0^2}
 - 
      \left| d_{2i} \right|^2
    \frac{ W_{i1} + W_{21} }
         {W_{i1}^2+ 2W_{i1} W_{21} + W_{21}^2 -\omega_0^2}
  \right\}
\\
&=&
      \frac{1}{ \hbar}
  \sum_i
  \left\{
      \left| d_{1i} \right|^2
    \frac{ W_{i1} }
         {W_{i1}^2-\omega_0^2}
 - 
      \left| d_{2i} \right|^2
    \frac{ W_{i1} + W_{21} }
         {
           \left( W_{i1}^2 -\omega_0^2 \right) 
         }
    \frac{ 1}
         {
           \left[ 
                  1 
                + \left( 2W_{21} / W_{i1} + W_{21}^2/ W_{i1}^2 \right)
                   W_{i1}^2 / \left( W_{i1}^2 -\omega_0^2 \right)
           \right]
         }
  \right\}
\\
&=&
      \frac{1}{ \hbar}
  \sum_i
  \left\{
      \left| d_{1i} \right|^2
    \frac{ W_{i1} }
         {W_{i1}^2-\omega_0^2}
 - 
      \left| d_{2i} \right|^2
    \frac{ W_{i1} + W_{21} }
         { W_{i1}^2 -\omega_0^2 }
           \left[ 
                  1 
                - \left( 2 \frac{W_{21}}{ W_{i1} } 
                         + \frac{W_{21}^2}{ W_{i1}^2}
                  \right)
                   \frac{W_{i1}^2}{ \left( W_{i1}^2 -\omega_0^2 \right)}
                + ...
           \right]
  \right\}
\\
&\approx&
      \frac{1}{ \hbar}
  \sum_i
  \left\{
      \left| d_{1i} \right|^2
    \frac{ W_{i1} }
         {W_{i1}^2-\omega_0^2}
 - 
      \left| d_{2i} \right|^2
    \frac{ W_{i1} + W_{21} }
         {W_{i1}^2 -\omega_0^2 }
           \left[ 
                  1 
                - 2 \frac{W_{21}}{ W_{i1} }
                    \frac{W_{i1}^2 }{ \left( W_{i1}^2 -\omega_0^2 \right)}
           \right]
  \right\}
\\
&\approx&
      \frac{1}{ \hbar}
 \sum_i
 \left\{
  \left[
      \left| d_{1i} \right|^2
 - 
      \left| d_{2i} \right|^2
 - 
      \left| d_{2i} \right|^2
      \frac{W_{21}}{ W_{i1}}
  \right]
    \frac{ W_{i1} }
         {W_{i1}^2-\omega_0^2}
 - 
  \left[
      2 \left| d_{2i} \right|^2
      \frac{W_{21}}{ W_{i1}}
  \right]
      \frac{ W_{i1}^3 }
           { \left( W_{i1}^2 -\omega_0^2 \right)^2 }
 \right\}
\label{eqn-alpha-stark-approx}
\end{eqnarray}

Ignoring the $W_{21}$ transition frequency terms, this is just the 
difference in the energy shifts of the two levels with 
field intensity.  
This is a purely {\em real} quantity, with no imaginary part.

% ---- ---- ----
\subsection{Two photon Bloch equations}

From now on, I use $\mathscr{I}$ (a script ``I'') to represent $E^2$ if
following on from \ref{sss-single-Efield} and eqns.(\ref{eqn-single-E-DcnDt}, 
\ref{eqn-single-E-alpha});
or to represent $2\left| A \right|^2$ if following on from 
\ref{sss-single-Eenvelope} and
eqns.(\ref{eqn-single-A-DcnDt}, \ref{eqn-single-A-alpha}). 
Since it is the 
envelope-carrier description of the field $E$ which is most useful 
(in \ref{sss-single-Eenvelope}), 
for most purposes $\mathscr{I} = 2\left| A \right|^2$ holds; 
remember that it is difficult to maintain the accuracy of the 
approximations relied on above for the $E^2$ picture 
(in \ref{sss-single-Efield}).
I also drop the prime on the $\alpha_{nj}'$ parameters used
 in the (Case Iii)) ``$2\left| A \right|^2$'' electric field envelope model.

Since we are only interested in $c_1, c_2$, and because we keep 
only stationary or slowly varying terms, we can write equations
for $c_1, c_2$ as --
~
\begin{eqnarray}
\imath \hbar
\frac{d c_1}{dt}
&=&
- \alpha_{11} \mathscr{I} c_1 
- \alpha_{12} \mathscr{I} c_2
\\
\Longrightarrow
~~~~ ~~~~
\imath \hbar
\frac{d c_1^*}{dt}
&=&
+ \alpha_{11}^* \mathscr{I} c_1^* 
+ \alpha_{12}^* \mathscr{I} c_2^*
\\
\imath \hbar
\frac{d c_2}{dt}
&=&
- \alpha_{21} \mathscr{I} c_1
- \alpha_{22} \mathscr{I} c_2
\\
\Longrightarrow
~~~~ ~~~~
\imath \hbar
\frac{d c_2^*}{dt}
&=&
+ \alpha_{21}^* \mathscr{I} c_1^*
+ \alpha_{22}^* \mathscr{I} c_2^*
\\
\textrm{In matrix form: } 
~~~~ ~~~~
\frac{d}{dt}
\left[
\begin {array}{c}
c_1
\\
c_2
\end {array}
\right]
&=&
  -\frac{1}{\imath \hbar}
 \left[
   \begin {array}{cc}
     \alpha_{11} \mathscr{I} & \alpha_{12} \mathscr{I} \\
     \noalign{\medskip}
     \alpha_{21} \mathscr{I} & \alpha_{22} \mathscr{I} \\
   \end {array}
  \right]
~
\left[
\begin {array}{c}
c_1
\\
c_2
\end {array}
\right]
\label{eqn-single-dcdt}
,
\end{eqnarray}

% ---- ---- ----
\subsection{Alternative formulations: Kien et.al. (KLKOHS), 
Hickman et.al. (HPB)}

This interlude compares my matrix equation for $c_1, c_2$ to those from 
Kien et.al.\cite{Kien-LKOHS-1999pra} (KLKOHS) and Hickman et.al. 
\cite{Hickman-PB-1986pra} (HPB).  This is useful as a fixed point of 
refernce beween the approaches, enabling quick conversions between the 
parameter variables.

\begin{eqnarray}
\textrm{cf. (KLKOHS) eqn.(13): } ~~~~ ~~~~
\frac{d}{dt}
\left[
\begin {array}{c}
c_a
\\
c_b
\end {array}
\right]
&=&
 -\frac{1}
       {\imath \hbar}
\left[
\begin {array}{cc}
\hbar \Omega_{aa} & \hbar \Omega_{ab} \\
\noalign{\medskip}
\hbar \Omega_{ba} & \hbar \Omega_{bb} - \delta \\
\end {array}
\right]
~
\left[
\begin {array}{c}
c_a
\\
c_b
\end {array}
\right]
,
\label{eqn-KLKOHS-dcdt}
\\
\textrm{where $B_q$ are KLKOHS's $E_q$ envelopes:}
~~~~ ~~~~
  E 
&=&
  \frac{1}{2} 
  \left[
     B_q e^{\imath \Xi} + B_q^* e^{-\imath \Xi} 
  \right]
\\
  \hbar \Omega_{aa}
&=&
  \frac{\hbar }{2}
  \sum_q
    a_q B_q B_q^*
\\
  \hbar \Omega_{ab}
&=&
  \frac{\hbar }{2}
  \sum_q
    d_q B_q B_q^*
~~~~ ~~~~
=
 \hbar \Omega_{ba}^*
\\
  \hbar \Omega_{bb}
&=&
  \frac{\hbar }{2}
  \sum_q
    b_q B_q B_{q+1}^* 
\end{eqnarray}

\begin{eqnarray}
\textrm{cf. (HPB) eqn.(13): } ~~~~ ~~~~
  \frac{d}{dt}
  \left[
   \begin {array}{c}
      c_1
      \\
      c_2
    \end {array}
  \right]
&=&
  -\frac{1}{\imath \hbar}
  \left[
    \begin {array}{cc}
      -H_{11} & -H_{12}  \\
      \noalign{\medskip}
      -H_{21} & -H_{22}  \\
    \end {array}
  \right]
~
\left[
\begin {array}{c}
c_1
\\
c_2
\end {array}
\right]
\label{eqn-HPB-dcdt}
,
\\
\textrm{where $\alpha_{Hij}$ are HPB's $\alpha_{ij}$ parameters:}
~~~~ ~~~~
  H_{11} 
&=&
  -\frac{1}{4}
 \sum_j \alpha_{H11} (\omega_j) V_j V_j^*
\\
  H_{12} 
&=&
  -\frac{1}{4}
 \sum_j \alpha_{H12} (\omega_j) V_j V_{j-1}^*
~~~~ ~~~~ = 
  H_{21}^*
\\
  H_{22} 
&=&
  -\frac{1}{4}
 \sum_j \alpha_{H22} (\omega_j) V_j V_j^*
\end{eqnarray}

Thus comparing the Rabi-like parts of my eqn (\ref{eqn-single-dcdt}) 
with that of KLKOHS (my eqn (\ref{eqn-KLKOHS-dcdt})) and that of 
HPB (my eqn (\ref{eqn-HPB-dcdt})), gives
~
\begin{eqnarray}
  \alpha_{12} \mathscr{I} = \alpha_{12} . 2 A^*A = \hbar f' . 2 A^*A
~~~~ ~~~~
&=&
  \frac{\hbar}{2}
  \sum_q
    d_q B_q B_{q+1}^*
=
  \hbar
  \sum_q
    d_q 
    .
    \frac{1}{2}
    B_q B_{q+1}^*
~~~~ ~~~~
= 
 -\frac{1}{4}
  \sum_j
    \alpha_{H12}(\omega_j) V_j V_{j-1}^*
\end{eqnarray}

\noindent
\textbullet ~~
This uses KLKOHS: 
$ E = \frac{1}{2} \left( B_1 + B_1^* + B_2 + B_2^* \right)$ \\
$\rightarrow E^2 = \frac{1}{4} \left( B_1^2 + B_1^{*2} + B_2^2 + B_2^{2*} 
+ 2 B_1 B_1^* + 2 B_2 B_2^* + 2 B_1 B_2^* + 2 B_1^* B_2 
\right)$ \\
$\rightarrow E_{1-2}^2 \sim \frac{1}{4} 2 B_1 B_2^* = \frac{1}{2} B_1 B_2^*$ \\
which tells us the size of the field contribution to the KLKOHS Rabi-like term.

\noindent
\textbullet  ~~
This uses HPB: 
which tells us the size of the field contribution to the HPB Rabi-like term.
(cf KLKOHS).
So $\Omega_{aa}=\alpha_{11}
\mathscr{I} / \hbar$, $\Omega_{ab}=\alpha_{12} \mathscr{I} / \hbar$,
$\Omega_{bb}= \alpha_{22} \mathscr{I} / \hbar$ (temporarily ignoring their
detuning $\delta$); and looking ahead to the definition of $f=
\bar{\alpha}_{12}/2\hbar =  \left( \alpha_{12}+\alpha_{21} \right) /2\hbar$
gives us $\Omega_{ab}=f \mathscr{I}$, noting $\mathscr{I}=2 A_q A_{q+1}^*$,
compared to $\mathscr{I}=\frac{1}{2}E_q E_{q+1}^*$.  Note that KLKOHS  have
defined $\Omega_{ab}=\frac{1}{2} \sum_q d_q E_q E_{q+1}^*$, so that $f=d_q$.

% ---- ---- ----
\subsection{Two photon Bloch equations (cont)}

An important difference between KLKOHS \& HPB, and my equations is 
that 
my $\alpha_{ij}$ coupling parameters retain a ``slow'' time dependence
at the Raman frequency $W_{12}$.  
I now turn my equations (\ref{eqn-single-dcdt}) into Bloch equations by 
first working out 
$d/dt$ 
of $c_1^* c_1$, $c_2^* c_2$, and $c_1^* c_2$, 
~
\begin{eqnarray}
 \imath \hbar
\frac{d c_1^* c_1}{dt}
=
 \imath \hbar
c_1^* \frac{d c_1}{dt}
+
 \imath \hbar
c_1 \frac{d c_1^*}{dt}
&=&
- \alpha_{11}   \mathscr{I} c_1^* c_1 
- \alpha_{12}   \mathscr{I} c_1^* c_2
+ \alpha_{11}^* \mathscr{I} c_1 c_1^*
+ \alpha_{12}^* \mathscr{I} c_1 c_2^*
\\
&=&
+ \alpha_{12}^* \mathscr{I} c_1 c_2^*
- \alpha_{12}   \mathscr{I} c_1^* c_2
- \left[ \alpha_{11} - \alpha_{11}^* \right] \mathscr{I} c_1^* c_1
\\
&=&
+ \mathscr{I} \left( \alpha_{12}^* c_1 c_2^* - \alpha_{12} c_1^* c_2 \right)
; ~~~~ ~~~~ ~~~~ ~~~~ \textrm{since} ~~ \alpha_{11}=\alpha_{11}^*
\\
 \imath \hbar
\frac{d c_2^* c_2}{dt}
=
 \imath \hbar
 c_2^* \frac{d c_2}{dt}
+
 \imath \hbar
 c_2 \frac{d c_2^*}{dt}
&=&
- \alpha_{21} \mathscr{I} c_2^* c_1 
- \alpha_{22} \mathscr{I} c_2^* c_2
+ \alpha_{21}^* \mathscr{I} c_2 c_1^*
+ \alpha_{22}^* \mathscr{I} c_2 c_2^*
\\
&=&
- \alpha_{21} \mathscr{I} c_2^* c_1 
+ \alpha_{21}^* \mathscr{I} c_2 c_1^*
- \left[ \alpha_{22} - \alpha_{22}^* \right] \mathscr{I} c_2^* c_2
\\
&=&
- \mathscr{I} \left( \alpha_{21}  c_2^* c_1 - \alpha_{21}^* c_2 c_1^* \right)
; ~~~~ ~~~~ ~~~~ ~~~~ \textrm{since} ~~ \alpha_{22}=\alpha_{22}^*
\\
 \imath \hbar
\frac{d c_1 c_2^*}{dt}
=
 \imath \hbar
c_1 \frac{d c_2^*}{dt}
+
 \imath \hbar
c_2^* \frac{d c_1}{dt}
&=&
+ \alpha_{21}^* \mathscr{I} c_1 c_1^*
+ \alpha_{22}^* \mathscr{I} c_1 c_2^*
- \alpha_{11} \mathscr{I} c_2^* c_1
- \alpha_{12} \mathscr{I} c_2^* c_2
\\
&=&
-\left( \alpha_{11} - \alpha_{22}^* \right) \mathscr{I} c_2^* c_1 
+ \alpha_{21}^* \mathscr{I} c_1 c_1^*
- \alpha_{12} \mathscr{I} c_2^* c_2
\\
&=&
- \left( \alpha_{11} - \alpha_{22}^* \right) \mathscr{I} c_1 c_2^* 
- \alpha_{12} \mathscr{I}  \left[ c_2 c_2^* - c_1^* c_1 \right]
; ~~~~ ~~~~ \textrm{since} ~~ \alpha_{21}^* \approx \alpha_{12}
\end{eqnarray}

Use $\rho_{12} = c_1 c_2^*$ %(NB hence the c.c. of $c_1^* c_2$) 
and $w=c_2 c_2^*-c_1^* c_1$,
~
\begin{eqnarray}
 \imath \hbar
\frac{d w}{dt}
&=&
- \mathscr{I} \left( \alpha_{21}   c_2^* c_1 - \alpha_{21}^* c_2 c_1^* \right)
- \mathscr{I} \left( \alpha_{12}^* c_1 c_2^* - \alpha_{12}   c_1^* c_2 \right)
\\
&=&
  \mathscr{I} 
  \left( 
   - \alpha_{12}^* c_1   c_2^* 
   + \alpha_{21}^* c_1^* c_2 
   - \alpha_{21}   c_1   c_2^*
   + \alpha_{12}   c_1^* c_2 
  \right)
\\
&=&
 -
  \mathscr{I} \left( \alpha_{12}^* + \alpha_{21}   \right) c_1 c_2^*
 +
  \mathscr{I} \left( \alpha_{21}^* + \alpha_{12}   \right) c_1^* c_2
\\
&=&
 -
  \mathscr{I} \left( \alpha_{12}^* + \alpha_{21}   \right) \rho_{12}
 +
  \mathscr{I} \left( \alpha_{21}^* + \alpha_{12}   \right) \rho_{12}^*
\\
 \imath \hbar
\frac{d \rho_{12}}{dt}
&=&
-
  \left( \alpha_{11} - \alpha_{22}^* \right) \mathscr{I} 
  c_1 c_2^*
- 
  \alpha_{12} \mathscr{I} 
  \left[ c_2 c_2^* - c_1^* c_1 \right]
\\
&=&
-
  \left( \alpha_{11} - \alpha_{22}^* \right) \mathscr{I} \rho_{12}
-
  \alpha_{12} \mathscr{I} w
\end{eqnarray}

Hence~
\begin{eqnarray}
\frac{d \rho_{12}}{dt}
&=&
  \imath 
  \frac{\left( \alpha_{11} - \alpha_{22}^* \right) }
     {\hbar}
  \mathscr{I} \rho_{12}
+ \imath 
  \frac{\alpha_{12} }{ \hbar}
  \mathscr{I} w
\label{eqn-basic2pbloch-rho}
\\
\frac{dw}{dt}
&=&
+ \imath
  \frac{\left( \alpha_{12}^* + \alpha_{21} \right) }
       { \hbar}
  \mathscr{I} 
    \rho_{12}
   - 
 \imath
  \frac{\left( \alpha_{21}^* + \alpha_{12} \right) }
       { \hbar}
  \mathscr{I} 
    \rho_{12}^*
,
\label{eqn-basic2pbloch-w}
\\
&=&
+ \imath
  \frac{2 \alpha_{12}^* }
       { \hbar}
  \mathscr{I} 
    \rho_{12}
   - 
 \imath
  \frac{ 2 \alpha_{12} }
       { \hbar}
  \mathscr{I} 
    \rho_{12}^*
.
\end{eqnarray}

% ---------------------------------
\subsection{Transformations of the Bloch Equations}\label{aa-single-transforms}

I define a new coupling parameter $f'$ (c.f. $\omega_B$), 
following the definition of $\bar{\alpha}_{12}$ in eqn.(\ref{eqn-alpha-bar}),
and
$\omega_b = W_2 - W_1 = W_{21}$.  I also define a $\omega_A$, 
which corresponds to the intensity dependent shift detuning 
shift proportional to  $(\alpha_{22} - \alpha_{11})$.
 Thus, factoring any complex phase of $\bar{\alpha}_{12}$ into 
the angle $\delta'$, I get --
~
\begin{eqnarray}
  f' e^{-\imath \omega_b t}
&=&
  \frac{\alpha_{12}}
       { \hbar}
~~~~ ~~~~
\simeq
  \frac{\bar{\alpha}_{12}}
       { 2 \hbar}  
  e^{-\imath \omega_b t - \imath \delta'}
; ~~~~ ~~~~
  \alpha_{12}
= 
  \frac{1}{\hbar} 
  e^{-\imath \omega_b t} 
  \sum_j d_{ij}d_{j2} \frac{-W_{j2}}{W_{j2}^2-\omega_0^2}
\label{eqn-blochcoupling}
\\
  g' 
&=&
  \frac{\alpha_{11} - \alpha_{22}^*}
       { \hbar}
~~~~ ~~~~
\textrm{NB: this is a real quantity, see eqns.
(\ref{eqn-alpha-stark}, \ref{eqn-alpha-stark-approx})}
\label{eqn-blochstark}
\\
\omega_B 
&=& \frac{\left( \alpha_{12} + \alpha_{21}^* \right) }
       { 2 \hbar}
  \mathscr{I} 
~~~~
= \frac{\bar{\alpha}_{12}}
       { 2 \hbar}
  e^{-\imath \omega_b t - \imath \delta'}
  \mathscr{I} 
~~~~
= 
  f' 
  e^{-\imath \omega_b t - \imath \delta'}
 \mathscr{I}
\label{eqn-blochcoupling-wB}
\\
\omega_A 
&=& 
  \frac{\left( \alpha_{11} - \alpha_{22}^* \right) }
       {\hbar}
   \mathscr{I}
\label{eqn-blochcoupling-wA}
.
\end{eqnarray}

The Bloch equations (\ref{eqn-basic2pbloch-rho}, \ref{eqn-basic2pbloch-w}) 
can now be rewritten --
~
\begin{eqnarray}
\frac{d \rho_{12}}{dt}
&=&
  \imath g' \mathscr{I} 
  \rho_{12}
+ 
  \imath f' \mathscr{I} w 
  e^{-\imath \omega_b t  - \imath \delta'}
\\
\frac{dw}{dt}
&=&
+
  2 \imath f' 
  \mathscr{I} 
  \left[ 
    \rho_{12}
    e^{+\imath \omega_b t  + \imath \delta'}
   - 
    \rho_{12}^*
    e^{-\imath \omega_b t  - \imath \delta'}
  \right]
.
\end{eqnarray}

In analogy to both the 
standard two level atom Bloch equations,
%(cf \href{../_pulse-fewcyc/twolevelatom}{Two level 
%atoms and the few cycle regime}, sections (2.4, 2.5))
 and those in HPB (HPB 18), these equations can have 
losses introduced, in --
~
\begin{eqnarray}
\frac{d \rho_{12}}{dt}
&=&
-
  \gamma_2 \rho_{12}
+
  \imath g' \mathscr{I} 
  \rho_{12}
+
  \imath f' \mathscr{I} w 
  e^{-\imath \omega_b t  - \imath \delta'}
\\
\frac{dw}{dt}
&=&
-
  \gamma_1
  \left(
    w - w_i
  \right)
+
  2 \imath f' 
  \mathscr{I} 
  \left[ 
    \rho_{12}
    e^{+\imath \omega_b t  + \imath \delta'}
   - 
    \rho_{12}^*
    e^{-\imath \omega_b t  - \imath \delta'}
  \right]
.
\end{eqnarray}

In the standard atom-field case,  the atom and field frequencies are similar,
so an atom carrier could be chosen to match its evolution the field carrier
frequency, leading naturally to a detuning term.  
However, in the Raman
situation, the atomic frequency is far removed from the field frequency, so 
the frequency evolution that appears in the definition of
$\psi$ (eqn.(\ref{eqn-psi-def})) is sufficient.

Here I define two things (a) new $u, v$ variables that represent the density
matrix element $\rho_{12}$, and (b) allow for a ``detuning'' rotation in
$\rho_{12}$. This detuning rotation looks rather like a carrier+envelope
representation for $\rho_{12}$, but without a ``+c.c.'' since $\rho_{12}$ is a
complex quantity.  There is no need for any kind of carrier+envelope
transformation for the inversion $w$.  

So,
~
\begin{eqnarray}
\rho_{12} 
= \frac{u}{2} + \imath \frac{v}{2}  
&=& 
\rho_{12}' \exp \left( -\imath \Delta t -\imath \delta' \right)
= \frac{u'}{2}  + \imath \frac{v'}{2}
\label{eqn-blochcoupling-rho12-uv}
\end{eqnarray}

%\newpage

I now adapt the Bloch equations to allow for the rotation in $\rho_{12}$,
and introduce a detuned transition frequency 
$\omega_b'=\omega_b-\Delta$,
~
\begin{eqnarray}
\partial_t 
\left( 
  \rho_{12}' e^{-\imath \Delta t -\imath \delta'} 
\right)
&=&
-
  \imath \Delta \rho_{12}
  e^{-\imath \Delta t-\imath \delta'}
+
  e^{-\imath \Delta t-\imath \delta'}
  \partial_t 
  \rho_{12}'
\\
&=&
-
  \gamma_2 
  \rho_{12}' e^{-\imath \Delta t-\imath \delta'} 
+
  \imath g' \mathscr{I} 
  \rho_{12}' e^{-\imath \Delta t-\imath \delta'} 
+ 
  \imath f' \mathscr{I}  w e^{-\imath \omega_b t-\imath \delta'} 
\\
\partial_t w
&=&
- 
  \gamma_1 \left( w - w_i \right)  
+
  2 \imath f' \mathscr{I}
  \left( 
    \rho_{12}'   
    e^{-\imath \Delta t-\imath \delta'} 
    e^{+\imath \omega_b t+\imath \delta'} 
   -
    \rho_{12}'^* 
    e^{+\imath \Delta t+\imath \delta'} 
    e^{-\imath \omega_b t-\imath \delta'} 
  \right)
\\
\Longrightarrow ~~~~ ~~~~
  e^{-\imath \Delta t-\imath \delta'}
  \partial_t 
  \rho_{12}' 
&=&
  \left(
     -\gamma_2 + \imath \Delta
  \right) 
  \rho_{12}' e^{-\imath \Delta t-\imath \delta'}
+
  \imath g' \mathscr{I} 
  \rho_{12}' e^{-\imath \Delta t-\imath \delta'}
+ 
 \imath f' \mathscr{I}  w e^{-\imath \omega_b t-\imath \delta'} 
\\
\partial_t w
&=&
- \gamma_1 \left( w - w_i \right)  
+
  2 \imath f' \mathscr{I}  
  \left( 
    \rho_{12}'
    e^{+\imath \omega_b' t} 
   -
    \rho_{12}'^*
    e^{-\imath \omega_b' t} 
  \right)
\\
\Longrightarrow ~~~~ ~~~~
  \partial_t \rho_{12}' 
&=&
  \left(
     -\gamma_2 + \imath \Delta
  \right) 
  \rho_{12}'
+
  \imath g' \mathscr{I} 
  \rho_{12}' 
+ 
    \imath f' \mathscr{I}  
    e^{\imath \omega_b' t} 
\\
\partial_t w
&=&
- \gamma_1 \left( w - w_i \right)  
+ 
  2 \imath f' \mathscr{I}  
  \left( 
    \rho_{12}'
    e^{+\imath \omega_b' t} 
   -
    \rho_{12}'^*
    e^{-\imath \omega_b' t} 
  \right)
\end{eqnarray}

Notice that we have made the fixed complex phase (the $\delta'$) vanish 
from the equations; this is not dependent on the presence of a finite
``detuning'' $\Delta$.  In what follows, $\Delta=0$, and there are
two cases depending on the chosen meaning for $\mathscr{I}$; however 
note that {\em both
give the same result}.

CASE (i): $\mathscr{I} = E^2$:  continue by applying the carrier+envelope
for the $E$ field $E = A e^{-\imath \omega_0 t} + A^* e^{+\imath \omega_0 t}$.

CASE (ii): $\mathscr{I} = 2A^*A$: we can jump 
straight to eqn.(\ref{eqn-rbpostRWA-w0-rho}, \ref{eqn-rbpostRWA-w0-w}) 
(i.e. $(D)$), since
the RWA was made when deriving the couplings $\alpha_{nj}$. 

The CASE (i) $\mathscr{I} = E^2$ starting point is --
~
\begin{eqnarray}
  \partial_t \rho_{12}'
&=&
  \left(
     -\gamma_2 + \imath \Delta + \imath g' \mathscr{I} 
  \right) 
  \rho_{12}'
 ~~~~
+ 
 \imath f' \mathscr{I} w
  e^{-\imath \omega_b' t} 
\\
\partial_t w
&=&
- \gamma_1 \left( w - w_i \right)  
 ~~~~
+  2 \imath f' \mathscr{I}
\left( 
  \rho_{12}'
  e^{+\imath \omega_b' t} 
-
  \rho_{12}'^*
  e^{-\imath \omega_b' t} 
\right)
,
\\
(B) \Longrightarrow ~~~~ ~~~~
  \partial_t \rho_{12}'
&=&
  \left(
     -\gamma_2 + \imath \Delta + \imath g' E^2
  \right) 
  \rho_{12}'
 ~~~~
\imath f' E^2 
    e^{-\imath \omega_b' t} 
  .
  w
\\
\partial_t w
&=&
- \gamma_1 \left( w - w_i \right)  
 ~~~~
+  2 \imath f' E^2
  \left( 
    \rho_{12}'
    e^{+\imath \omega_b' t } 
  -
    \rho_{12}'^* 
    e^{-\imath \omega_b' t } 
  \right)
,
\\
(C) \Longrightarrow ~~~~ ~~~~
  \partial_t \rho_{12}' 
&=&
  \left(
     -\gamma_2 + \imath \Delta + \imath g' E^2
  \right) 
  \rho_{12}'
 ~~~~
+ 
 \imath f' 
\left[ 
  A e^{-\imath \omega_0 t} + A^* e^{+\imath \omega_0 t} 
\right]^2
 w 
    e^{-\imath \omega_b' t} 
\\
\partial_t w
&=&
- 
 \gamma_1 \left( w - w_i \right)  
 ~~~~
+  
 2 \imath f' 
\left[ 
  A e^{-\imath \omega_0 t} + A^* e^{+\imath \omega_0 t} 
\right]^2
\left( 
  \rho_{12}
    e^{+\imath \omega_b' t} 
-
  \rho_{12}^*
    e^{-\imath \omega_b' t} 
\right)
,
~~~~
\\
(\textrm{apply a RWA about}~ \omega_0) ~~~~ ~~~~ ~~~~ ~~~~ 
&:::& 
\left[ 
  A e^{-\imath \omega_0 t} + A^* e^{+\imath \omega_0 t} 
\right]^2 
\nonumber
\\
&& ~~~~ ~~~~ =
\left[ 
  A^2   e^{-2 \imath \omega_0 t} 
+ 2 A^* A 
+ {A^*}^2 e^{+2 \imath \omega_0 t} 
\right]
\\
\Longrightarrow  ~~~~ ~~~~
&& ~~~~ ~~~~ \approx
2 A^* A 
\\
CASE(ii) ~\&~ (D) \Longrightarrow ~~~~ ~~~~
  \partial_t \rho_{12}' 
&=&
  \left(
     -\gamma_2 + \imath \Delta + 2 \imath g' A^* A
  \right) 
  \rho_{12}'
 ~~~~
+ 
2 \imath f' A^* A w
    e^{-\imath \omega_b' t} 
\label{eqn-rbpostRWA-w0-rho}
\\
\partial_t w
&=&
- 
  \gamma_1 \left( w - w_i \right)  
 ~~~~
+ 
  4\imath f' 
    A^* A 
\left( 
  \rho_{12}
    e^{+\imath \omega_b' t } 
-
  \rho_{12}^*
    e^{-\imath \omega_b' t } 
\right)
\label{eqn-rbpostRWA-w0-w}
,
\\
(E) ~~~~ ~~~~
  \partial_t \rho_{12}' 
&=&
  \left(
     -\gamma_2 + \imath \Delta + 2 \imath g' A^* A
  \right) 
  \rho_{12}'
+ 
  2 \imath f' A^*A w .
    e^{-\imath \omega_b' t} 
\label{eqn-rbpostRWA-wb-rho}
\\
\partial_t w
&=&
- 
  \gamma_1 \left( w - w_i \right)  
+ 
  4 \imath f' 
  A^* A
  \left( 
    \rho_{12}'
      e^{+\imath \omega_b' t}  
   -
    \rho_{12}'^* 
      e^{-\imath \omega_b' t} 
  \right) 
\label{eqn-rbpostRWA-wb-w}
,
\\
(F: \textrm{split ~} \rho_{12},) \Longrightarrow ~~~~ ~~~~
  \partial_t u'
&=&
-
  \gamma_2 u' 
- 
  \left( \Delta + 2 g' A^* A \right) v' 
  ~~~~
+ 
  4 f' A^* A w  . \sin \left( \omega_b' t \right)
\label{eqn-rbpostRWA-last-u}
\\
  \partial_t v'
&=&
-
  \gamma_2 v' 
+ 
  \left( \Delta + 2 g' A^* A \right) u'
  ~~~~
+ 
  4 f' A^* A w  . \cos \left( \omega_b' t \right)
\label{eqn-rbpostRWA-last-v}
\\
\partial_t w
&=&
- \gamma_1 \left( w' - w_i \right)  
~~~~
 -  4  f' A^* A u'  . \sin \left( \omega_b' t \right)
~~~~
 -  4  f' A^* A v'  . \cos \left( \omega_b' t \right)
,
\label{eqn-rbpostRWA-last-w}
\end{eqnarray}

Note $2\rho_{12} = u+\imath v$, since $\rho_{12}=c_1 c_2^*$, as
per eqn.(\ref{eqn-blochcoupling-rho12-uv}); this means the $2$ in the
$u,v$ equations becomes $4$, whereas the $4$ in the $w$ equation 
isn't doubled to $8$.  Note (again) that the Stark shift parameter 
$g'$ is real valued.

\subsubsection{Rotations}

Looking at eqns.(\ref{eqn-rbpostRWA-last-u}, \ref{eqn-rbpostRWA-last-v}, 
\ref{eqn-rbpostRWA-last-w}), 
 we can see three rotation angles: $\theta_{uv} = \Delta$, 
 $\theta_{uw} = 4 f' A^* A \sin \left( \omega_b t - \delta' \right) $, and 
 $\theta_{vw} = 4 f' A^* A \cos \left( \omega_b t - \delta' \right)$, 
 which apply to the coordinate pairs 
 $\left( u', v' \right) $, $\left( u', w\right) $, 
 and $\left( v', w\right) $ respectively. 
For extra generality, 
 I will allow a complex valued $f' = f_r + \imath f_i$,
 but note that previous definitions make $f'$ real.  )

In vector notation, ignoring the losses, we can construct a torque 
vector $\vec{\Omega}$ (unchecked signs), 
~
\begin{eqnarray}
\partial_t u' &=& 0.u' - \theta_{uv}.v' + \theta_{uw}.w
\\
\partial_t v' &=& \Delta.u' + 0.v' - \theta_{vw}.w
\\
\partial_t w &=& -\theta_{uw}.u' + \theta_{vw}.v' + 0.w
\\
\frac{d}{dt} \left[ u', v', w \right]
&=&
\left[ -4 f'_r A^* A,
       ~~ -4 f'_i A^* A,
       ~~ \Delta
\right]
\\
&=&
\left[ \theta_{vw}, ~ \theta_{uw}, ~ \theta_{uv}
\right]
\times
\left[ u', v', w \right] 
\\
\Longrightarrow ~~~~ ~~~~
\frac{d}{dt} \vec{U} &=& \vec{\Omega} \times \vec{U}
\end{eqnarray}

I should now be able to turn this into a (unitary) rotation matrix 
 for the $\vec{U}$ vector.

% (see twolevelatom.dvi).

%Also I need to derive the polarization term that will act back
%on the (propagating) $E$ field.

% ------------------------
\subsection{The polarization driving the field}\label{ss-single-driving}

There is a distinction between the atomic polarization 
 $P$, and the effect on the field of that atomic polarization.
This is beause we are dealing with a nonlinear interaction.
 I denote the (effective) Raman polarization $\mathscr{P}$, and this 
 quantity is the one that appears in the wave equation.
Allen and Eberly {\em ``Optical resonance and two level atoms''}
 \cite{AllenEberly-ORTLA} have,
 for the standard (non-Raman) case (skipping the integral), an 
 eqn.(AE 4.2) --
~
\begin{eqnarray}
(AE ~ 4.2) ~~~~ ~~~~ ~~~~ ~~~~
P(t,z) &=& 
\mathscr{N} d 
\left[ 
    u \cos(\omega t - Kz) - v \sin(\omega t - Kz)
\right]
\end{eqnarray}

HPB, after summing the electric field components (which absorbs
a $1/2$), have
~
\begin{eqnarray}
(HPB ~ 21) ~~~~ ~~~~ ~~~~ ~~~~
P &=& 
 \frac{1}{4}
 e^{-\imath \theta}
\left( 
  u + \imath v
\right)
\sigma
\alpha_{12}
\sum_j
\left[
  V_j e^{ \imath \omega_{j-1} t}
 +
  V_j^* e^{ -\imath \omega_{j+1} t}
\right]
~~~~ 
+ c.c.
\\
&=&
 \frac{1}{2}
 e^{-\imath \theta}
\left( 
  u + \imath v
\right)
\sigma
\alpha_{12}
  e^{ -\imath \omega_b t}
 \frac{1}{2}
\sum_j
\left[
  V_j e^{ \imath \omega_j t}
 +
  V_j^* e^{ -\imath \omega_j t}
\right]
~~~~ 
+ c.c.
\\
&=&
 \frac{1}{2}
 e^{-\imath \theta}
\left( 
  u + \imath v
\right)
\sigma
\alpha_{12}
  e^{ -\imath \omega_b t}
E
~~~~ 
+ c.c.
\end{eqnarray}

I now write down the the polarization envelope $B$, 
in my variables, and based on the same carrier as the electric field.
I also introduce a complex factor equivalent to HPB's $e^{ -\imath \theta}$,
for closer matching of the expressions.
%Any complex phase $\delta'$ from
%$\bar{\alpha}_{12}$ is added into the angle $\theta$ to give $\theta'$
%and (real) $\bar{\alpha}_{12}'$.
~
\begin{eqnarray}
\mathscr{P} ~~
= 
  \mathscr{B} e^{-\imath \Xi}
 +
  \mathscr{B}^* e^{+\imath \Xi}
 &=& 
 \rho_{12}
 e^{-\imath \theta }
\sigma
\alpha_{12}
\left( 
  A e^{-\imath \Xi}
 +
  A^* e^{+\imath \Xi}
\right)
~~~~
+ c.c.
\\
 &=& 
 \zeta \rho'_{12}
 e^{-\imath \theta}
\sigma . 
     \bar{\alpha}_{12}
           e^{-\imath \omega_b' t } 
\left( 
  A e^{-\imath \Xi}
 +
  A^* e^{+\imath \Xi}
\right)
~~~~ 
+ c.c.
\\
 &=& 
 \zeta 
 \rho'_{12}
 e^{-\imath \theta}
\sigma . 
     \bar{\alpha}_{12}'
           e^{-\imath \omega_b' t } 
\left( 
  A e^{-\imath \Xi}
 +
  A^* e^{+\imath \Xi}
\right)
+ 
 \zeta 
 {\rho'_{12}}^*
 e^{+\imath \theta}
\sigma . 
     \bar{\alpha}_{12}'
           e^{+\imath \omega_b' t } 
\left( 
  A^* e^{+\imath \Xi}
 +
  A e^{-\imath \Xi}
\right)
\\
&=&
 \zeta 
 \rho'_{12}
 e^{-\imath \theta}
\sigma . 
     \bar{\alpha}_{12}'
           e^{+\imath \omega_b' t } 
  A e^{-\imath \Xi}
~~~~
+
 \zeta 
 {\rho'_{12}}^*
 e^{+\imath \theta}
\sigma . 
     \bar{\alpha}_{12}'
           e^{-\imath \omega_b' t } 
  A e^{-\imath \Xi}
\nonumber
\\
&& + 
 \zeta 
 \rho'_{12}
 e^{-\imath \theta}
\sigma . 
     \bar{\alpha}_{12}'
          e^{+\imath \omega_b' t } 
  A^* e^{+\imath \Xi}
~~~~
+
 \zeta 
 {\rho'_{12}}^*
 e^{+\imath \theta}
\sigma . 
     \bar{\alpha}_{12}'
           e^{-\imath \omega_b' t } 
  A^* e^{+\imath \Xi}
\\
&=&
 \zeta 
 \sigma . 
 \bar{\alpha}_{12}'
\left\{
 \rho'_{12}
           e^{+\imath \omega_b' t -\imath \theta} 
+
 {\rho'_{12}}^*
           e^{-\imath \omega_b' t +\imath \theta} 
\right\}
  A e^{-\imath \Xi}
\nonumber
\\
&& + 
 \zeta 
 \sigma . 
 \bar{\alpha}_{12}'
\left\{
  \rho'_{12}
          e^{+\imath \omega_b' t -\imath \theta} 
+
 {\rho'_{12}}^*
           e^{-\imath \omega_b' t +\imath \theta} 
\right\}
  A^* e^{+\imath \Xi}
\\
&=&
 \zeta 
 \sigma . 
 \bar{\alpha}_{12}'
 X(t)
  A e^{-\imath \Xi}
 + 
 \zeta 
 \sigma . 
 \bar{\alpha}_{12}'
 X(t)
  A^* e^{+\imath \Xi}
\\
\textrm{ where the real valued is } ~~~~
  X(t) 
&=& 
  \left\{
    \rho'_{12}
           e^{+\imath \omega_b' t  -\imath \theta} 
    +
    {\rho'_{12}}^*
           e^{-\imath \omega_b' t  +\imath \theta}     
  \right\}
\label{eqn-single-Xrho}
\\
&=&
  \left\{
    \frac{1}{2}
    \left( u' + \imath v' \right)
           e^{+\imath \omega_b' t -\imath \theta} 
   +
    \frac{1}{2}
    \left( u' - \imath v' \right)
           e^{-\imath \omega_b' t +\imath \theta} 
  \right\}
\\
&=&
  \left\{
    u' \cos \left( \omega_b' t - \theta \right)
   -
    v' \sin  \left( \omega_b' t - \theta \right)
  \right\}
\label{eqn-single-X}
%  \frac{1}{2}
\\
\textrm{and} ~~~~ ~~~~
  \mathscr{B}(t) 
&=&
  \zeta 
  \sigma \bar{\alpha}_{12}'
  X(t) A(t)
\end{eqnarray}

From the post-$\omega_0$
RWA at eqns.(\ref{eqn-rbpostRWA-last-u}, 
\ref{eqn-rbpostRWA-last-v}, \ref{eqn-rbpostRWA-last-w}); above; and an 
un-scaled eqn. (\href{../_pulse-fewcyc/fewcyc.dvi}{FCPP} 3.48), --
~
\begin{eqnarray}
  \partial_t u'
&=&
  -\gamma_2 u' 
  -\left( \Delta + 2 g' A^* A \right) v' 
  ~~~~
  + 4 f' A^* A w \sin( \omega_b' t )
\label{eqn-single-Aprop-du}
\\
  \partial_t v'
&=&
  -\gamma_2 v 
  + \left( \Delta + 2 g' A^* A \right) u'
  ~~~~
  + 4 f' A^* A w \cos( \omega_b' t )
\label{eqn-single-Aprop-dv}
\\
\partial_t w
&=&
- \gamma_1 \left( w' - w_i \right)  
  ~~~~
-  4 f' A^* A . \sin(\omega_b' t ) u'
  ~~~~
-  4 f' A^* A . \cos(\omega_b' t ) v'
,
\label{eqn-single-Aprop-dw}
\\
\partial_z A(t) 
&=&
\frac{2 \imath \pi \omega_0}{c_0 n_0} 
\frac{ \mathscr{B}(t) }{4 \pi \epsilon_0}
,
~~~~ ~~~~
\textrm{ NB: units } [A]/m=\frac{s^{-1}}{m.s^{-1}} [\mathscr{B}]
~~\rightarrow [A] = [\chi] [B] [A]
~~\rightarrow [\chi] = [B^{-1}] = 1
\\
&=&
  \frac{\imath \omega_0}{2 c_0 n_0 \epsilon_0} 
  \mathscr{B}(t)
\\
&=&
\imath 
\frac{\zeta \sigma \bar{\alpha}_{12} \omega_0}{2 c_0 n_0 \epsilon_0 } 
 ~.~
    \left[ 
       u' \cos \left( \omega_b' t - \theta \right)
      -
       v' \sin \left( \omega_b' t  - \theta \right)
    \right]
 ~.~
 A(t) 
\\
&=&
\imath 
\frac{\zeta \sigma \bar{\alpha}_{12} \omega_0}{2 c_0 n_0 \epsilon_0 } 
 ~.~
A(t) X(t)
\label{eqn-single-Apropagate}
\end{eqnarray}

Comparing the prefactor of eqn.(\ref{eqn-single-Apropagate}) to that of 
the corrected (HPB 22), 
~
\begin{eqnarray}
\textrm{(HPB 22)} ~~~~ ~~~~
\frac{\partial V_j}{\partial z}
&=&
\frac{\sigma \alpha_{12}}{4 \epsilon_0 c}
\omega_j
\left[
  e^{-\imath \theta} \left(u-\imath v\right) V_{j+1}
 -
  e^{+\imath \theta} \left(u+\imath v\right) V_{j-1}
\right]
\end{eqnarray}
~
we can see that the only apparent differences are a factor of $n_0$, 
and in that HPB have a $e^{\pm \imath \theta}$ term that I omit.  By 
imagining the
cosine term split up into $+$ and $-$ frequency exponentials, 
it is easy to see how the relations between $V_j$ and $V_{j\pm 1}$ 
will arise.
Note the appearance (in HPB) of a carrier-dependent $\omega_i$ term, which 
in a standard multi-field approach would lead to different prefactors on 
the different $\partial_\xi A_i$ equations.

\subsubsection{Simulation ``photon'' variables}

Scale the field $E$ into square-root intensity ``photon'' variables, from 
$I = 2 n_0 c_0 \epsilon_0 E^2$, so that 
~
\begin{eqnarray}
A_p(t)
&=&
\sqrt{2 c_0 n_0 \epsilon_0 } ~ . ~ A(t)
\label{eqn-aphotonscale}
\\
f_p &=& 
\frac{f'}{2 c_0 n_0 \epsilon_0};
~~~~ ~~~~
g_p ~=~
\frac{g'}{2 c_0 n_0 \epsilon_0};
\label{eqn-fphotonscale}
\end{eqnarray}

Hence $f' A^*A \longrightarrow f_p A_p^* A_p$, so that
~
\begin{eqnarray}
  \partial_t u'
&=&
-
  \gamma_2 u' 
-
  \left( \Delta + 2 g_p A^* A \right) v' 
  ~~~~
+ 
  4 f_p A_p^* A_p w \sin(\omega_b' t)
\\
  \partial_t v'
&=&
- 
  \gamma_2 v 
+ 
   \left( \Delta + 2 g_p A^* A \right) u'
  ~~~~
+ 
   4 f_p A_p^* A_p w \cos(\omega_b' t)
\\
\partial_t w
&=&
- \gamma_1 \left( w' - w_i \right)  
  ~~~~
-  4 f_p A_p^* A_p u'  \sin(\omega_b' t)
  ~~~~
-  4 f_p A_p^* A_p v'  \cos(\omega_b' t)
\\
  \partial_z A_p(t)
&=&
  \imath
  \frac{\zeta \sigma \bar{\alpha}_{12} \omega_0}
       { 2 c_0 n_0 \epsilon_0}
  A_p(t)
  X(t)
\label{eqn-single-Appropagate2}
\\
&=&
  \imath
  \zeta \sigma \omega_0
  \frac{ 4 \hbar c_n n_0 \epsilon_0 f_p}
       { 2 c_0 n_0 \epsilon_0}
  A_p(t)
  X(t)
\\
&=&
  \imath
  \frac{\zeta}{2} 
  \left( 4 \sigma \hbar \right)
  \omega_0 f_p
  A_p(t)
  X(t)
~~~~ ~~~~
=
  \imath
  \frac{\zeta}{2}
  \omega_0 R f_p
  A_p(t)
  X(t); ~~~~
  R = 4 \sigma \hbar
\label{eqn-single-Apropphoton}
\end{eqnarray}

since 
~
\begin{eqnarray}
  \bar{\alpha}_{12} = 2 \hbar f' 
~~~~ \rightarrow 
  \frac{\bar{\alpha}_{12}}{2 c_0 n_0 \epsilon_0} = 2 \hbar f_p
~~~~ \rightarrow 
  \bar{\alpha}_{12} = 4 \hbar c_0 n_0 \epsilon_0  f_p
\end{eqnarray}

\noindent
Scalings: \\
$\hbar = 1.05\times10^{-34}$Js $\rightarrow 1.05\times10^{-10}$nJ.fs\\
$\sigma = X$m$^{-3} \rightarrow X\times10^{-18} \mu$m$^{-3}$\\
$2 c_0 n_0 \epsilon_0 = 5.31 \times 10^{-3}$m.s$^{-1}$~.~J.m$^{-1}$.V$^{-2}
\rightarrow 5.31 \times 10^{-3}$~.~J.s$^{-1}$.V$^{-2}
\rightarrow 5.31 \times 10^{-9}$~.~nJ.fs$^{-1}$.V$^{-2}$

\noindent
Scalings: \\
$[c \epsilon_0 ] = m.s^{-1} ~.~ J.m^{-1}.V^{-2} ~~~ = J.s^{-1}.C^2.J^{-2}
~~~ = C^2. J^{-1}.s^{-1} ~~~ = C^2 ~.~ 10^{-9} nJ^{-1} ~.~ 10^{-15} fs^{-1}
~~~ = 10^{-24} . C^2 .  nJ^{-1} . fs^{-1} $\\
$[\epsilon_0 ] = J.m^{-1}.V^{-2} ~~~ = J.m^{-1}.C^2.J^{-2}
~~~ = C^2 . J^{-1} . m^{-1} = C^2 ~.~ 10^{-9} nJ^{-1} ~.~ 10^{-6} \mu m^{-1}
~~~ = 10^{-15} . C^2 .  nJ^{-1} . \mu m^{-1} $\\

\subsubsection{Comments on the field propagation}

We can  see that the 
derivative of $A$ is proportional to $\imath A$, hence the 
(time-domain) evolution just rotates each point $A(t)$ 
differently without changing its amplitude. This might 
seem to imply that we will never get a shorter pulse than
we put in; but note that the $A$ is an {\em envelope}, and the
field is $A + A^*$, so that in principle the phase of $A$ might
be  such
that its amplitudes cancel in certain $t$ regions but not others, 
leading to a shorter pulse (and the magnetic field $H$ also, since
the fields are plane polarized).  
However, it seems unlikely that this will 
happen (barring some miraculous coincidence) from purely 
Raman effects.

If I can predict the output spectral phases though, a structure with a
suitably designed dispersion {\em might} be able to impose the desired
phases.  It would only be necessary to get the dispersion right at the comb
points of the spectrum.  See Shverdin et.al. \cite{Shverdin-WYYH-2004pre}, 
who do a four-wave mixing optimization procedure in their experiment to match
their phases appropriately.

% ------------------------------
\subsection{The steady state and gain co-efficient}\label{ss-steadystate}

Starting from the eqns.(\ref{eqn-single-Aprop-du}, 
\ref{eqn-single-Aprop-dv}, \ref{eqn-single-Aprop-dw}),
assume $v'$ is steady state (NB $\gamma_2 = 1/T_2$), so (with $w=-1$) -- 
~
\begin{eqnarray}
\partial_t \approx 0
~~~~ ~~~~
&=& 
-\gamma_2 v' + 4 f' A^* A  (-1) \cos(\omega_b t)
\\
\Longrightarrow ~~~~ ~~~~
v_0' &=& 
\frac{4 f'}{\gamma_2}  ~.~  A^* A . \cos(\omega_b t)
~~~~ ~~~~ =
4 f' T_2 ~.~  A^* A . \cos(\omega_b t)
\end{eqnarray}

I now use the field propagation equation (\ref{eqn-single-Apropphoton}), 
with the $A$ field envelope 
on each side scaled into ``photon'' variables.  Inserting the 
steady state of $v'$ calculated immediately above, we have --
~
\begin{eqnarray}
\partial_z A_p 
&=& 
\imath
\frac{\zeta \sigma \bar{\alpha}_{12} \omega_0}
     {2 c_0 n_0 \epsilon_0}
A_p v_0' \sin(\omega_b t)
\\
&=& 
\imath
\frac{\zeta \sigma \bar{\alpha}_{12} \omega_0}
     { 2 c_0 n_0 \epsilon_0}
A_p 
~.~
4 f' T_2 ~.~  A^* A . \cos(\omega_b t) .  \sin(\omega_b t)
\\
&=& 
\imath
\frac{\zeta \sigma \bar{\alpha}_{12} \omega_0}
     {2 c_0 n_0 \epsilon_0}
A_p 
~.~
  4 T_2
\frac{\bar{\alpha}_{12}}
     {2 \hbar} 
\frac{A_p^* A_p}
     {2 c_0 n_0 \epsilon_0} 
   \left[ \cos(\omega_b t) .  \sin(\omega_b t) \right]
\\
&=& 
\imath
\frac{ \zeta \sigma \omega_0 T_2 \bar{\alpha}_{12}^2}
     { 2 c_0^2 n_0^2 \epsilon_0^2 \hbar }
A_p 
  A_p^* A_p .
   \left[ \cos(\omega_b t) .  \sin(\omega_b t) \right]
\\
&=& 
 \imath G_p 
A_p 
  A_p^* A_p .
   \left[ \cos(\omega_b t) .  \sin(\omega_b t) \right]
,
\\
G_p &=& 
\frac{ \zeta }{2}
\frac{ \sigma \omega_0 T_2 \bar{\alpha}_{12}^2}
     { c_0^2 n_0^2 \epsilon_0^2 \hbar }
\label{eqn-ssgain-Gp}
.
\end{eqnarray}

Apparently, therefore, 
this $G_p$ corresponds to the usual ``gain co-efficient'' $g$:
~
\begin{eqnarray}
\textrm{cf (SMN) } ~~~~ ~~~~
g' &=&
\frac{ \sigma \omega_0 T_2 \alpha_{12}^2 }
     { c^2 \epsilon_0^2 \hbar }
\label{eqn-ssgain-gp-smn}
,
\end{eqnarray}

and we can assume that they are identical but for a factor of $2 n_0^2 /
\zeta$.  I do not know whether SMN silently assumes $n_0=1$, or factors
$n_0$ into $c$. In any case $n_0=1$ is usually accurate enough in gases.  

However, one hidden
complication with my above equation is that $A_p$ includes both the 
center ``pump'' field and the Raman sideband we are amplifying -- thus to 
properly check the gain co-efficient, we should expand it into its 
components, as is done below.

% ------------------------------
\subsubsection{Pump and sideband calculation}

$A_p$ includes both fundamental $A_1$ and its Raman sideband $A_2=A_2' 
e^{-\imath \omega_b t}$.  For the moment, I leave the calculation in a  rather 
abbreviated state -- probably it should be shifted to the Multi-field 
section following later. So
~
\begin{eqnarray}
A_p A_p^* A_p
&=&
 \left[ A_1   + A_2   \right]
 \left[ A_1^* + A_2^* \right]
 \left[ A_1   + A_2   \right]
\\
&=&
  \left[ 
    A_1 A_1^* + A_1 A_2^* + A_2 A_1^* + A_2 A_2^*
  \right]
 \left[ A_1   + A_2   \right]
\\
&=&
    A_1 A_1^* A_1 + A_1 A_2^* A_1 + A_2 A_1^* A_1 + A_2 A_2^* A_1 
 +
    A_1 A_1^* A_2 + A_1 A_2^* A_2 + A_2 A_1^* A_2 + A_2 A_2^* A_2
\\
\textrm{1st order terms in $A_2$ only:}
~~~~ ~~~~
&\simeq&
  A_1^2 A_2^* + 2 A_2 A_1^* A_1
\\
\textrm{drop counter rotating $A_1^2 A_2^*$: }
~~~~ ~~~~
&\simeq&
  2 A_2 A_1^* A_1
.
\end{eqnarray}

Thus the sideband envelope $A_2'$ evolves as 
~
\begin{eqnarray}
\partial_z A_2' 
&=& 
 \imath G_p 
 A_2' 
 .
  2 A_1^* A_1 . \cos(\omega_b t)
\\
&=& 
 \imath G_p 
   A_2'
 .
   I_1 . \cos(\omega_b t)
\\
&=& 
 \imath G'_p 
  A_2'
 .
  I_1 . \cos(\omega_b t)
,
\\
\textrm{where}
~~~~ ~~~~
  G'_p
&=& 
  \frac{ \zeta }{2}
  \frac{ \sigma \omega_0 T_2 \bar{\alpha}_{12}^2}
       { c_0^2 n_0^2 \epsilon_0^2 \hbar }
~~~~ ~~~~
=
  G_p
\end{eqnarray}

since $I_m = \frac{1}{2} A_m^* A_m$.  I don't convert $A_2'$ into 
$I_2$ because it occurs equally on both side of the equation.

% ----------------------------------------------------------------------
%\newpage
\section{Multi-field variant of single-field Raman theory}
\label{s-multifield}

The single-field Raman model above can be converted into a  traditional
multi-field model as developed in e.g.  HPB \cite{Hickman-PB-1986pra} or Syed,
McDonald and New 
\cite{Syed-MN-2000josab} by  replacing the field envelope with a sum of
multiple envelopes using carrier exponentials spaced at the Raman frequency.
When doing this, I will only get the correct multi-field form if few-cycle 
(either SEWA or GFEA) corrections to the field evolution part of
the theory are applied to the effective polarization caused by the 
Raman transition.

The idea is to replace the single field envelope with a sum of multiple 
envelopes spaced at the Raman frequency, which are best placed to 
represent the comb of frequencies generated by the Raman interaction.
Note that it will
not necessarily be identical to HPB and/or SMN, 
because the field equations are derived
from a propagation equation using a $\omega_0, \beta_0$ carrier, but it will
be very closely related.

Starting from eqns.(\ref{eqn-rbpostRWA-wb-rho},\ref{eqn-rbpostRWA-wb-w}),
Since the single-field evolution equation (eqn.(\ref{eqn-single-Apropagate})) 
uses an envelope $A$ that is based on a carrier
(see eqn.(\ref{eqn-single-EfromA})), 
the single-field envelope $A$ is replaced with $A_j$'s 
at frequency $\omega_j = \omega_0 + j \omega_b$ with wavevector $k_j =
k(\omega_j)$; thus $\omega'_j = \omega_j - \omega_0$, 
$k'_j = k_j - k(\omega_0) = k_j - k_0$;
also $\beta \leftrightarrow k$.
The single-field envelope in terms of the new $A_j$'s is
~
\begin{eqnarray}
A = \sum_j A_j \exp\left[ -\imath \left( \omega_j t - k_j z \right) \right]
\end{eqnarray}

% ------------------------------
\subsection{Polarization ($ \rho_{12}'$)}\label{ss-multifield-rho}

First I will handle the polarization ($ \rho_{12}'$) equation (from 
eqn.(\ref{eqn-rbpostRWA-w0-rho}))  (watch for any 
$\omega_b' = \omega_b - \Delta$ 
confusion, and note $2\rho_{12}'=u'+\imath v'$) -- 
~
\begin{eqnarray}
  \partial_t \rho_{12}' 
&=&
  \left(
     -\gamma_2 + \imath \Delta + 2 \imath g' A^* A
  \right) 
  \rho_{12}'
+ 
  2 \imath f' A^*A w 
     e^{-\imath \omega_b' t } 
\\
(A1) ~~ \Longrightarrow ~~
  \partial_t \rho_{12}' 
&=&
  \left(
     -\gamma_2 + \imath \Delta 
  \right) 
  \rho_{12}'
+
  2 \imath g' \rho_{12}'
    \sum_j \sum_k A_j^*A_k 
      e^{+\imath \left( \omega_j - \omega_k \right) t} 
      e^{+\imath \left( -k_j + k_k \right) z } 
\nonumber
\\
&& ~~~~ ~~~~
+ 
2 \imath f' \sum_j \sum_k A_j^*A_k w 
e^{+\imath \left( \omega_j - \omega_k - \omega_b' \right) t} 
e^{+\imath \left( -k_j + k_k \right) z } 
\\
&=&
  \left(
     -\gamma_2 + \imath \Delta 
  \right) 
  \rho_{12}'
+
  2 \imath g' \rho_{12}'
    \sum_j \sum_k A_j^*A_k 
      e^{+\imath \left( j - k \right) \omega_b t} 
      e^{+\imath \left( -k_j + k_k \right) z } 
\nonumber
\\
&& ~~~~ ~~~~
+ 
2 \imath f' \sum_j \sum_k A_j^*A_k w 
e^{+\imath \left( j - 1 - k \right) \omega_b t + \imath \Delta t} 
e^{+\imath \left( k_k-k_j  \right) z } 
\\(RWA) ~~~~ ~~~~
&\approx&
  \left(
     -\gamma_2 + \imath \Delta 
  \right) 
  \rho_{12}'
+
  2 \imath g' \rho_{12}'
    \sum_j A_j^*A_j w 
%\nonumber
%\\
%&& ~~~~ ~~~~
+ 
4 \imath f' \sum_j A_j^*A_{j-1} 
. w 
. e^{+\imath \Delta t } 
. e^{+\imath \left( k_{j+1}-k_j  \right) z } 
\\
&\approx&
  \left(
     -\gamma_2 + \imath \Delta + 2 \imath g' \sum_j A_j^* A_j
  \right) 
  \rho_{12}'
+ 
4 \imath f' \sum_j A_{j}  A_{j+1}^*
. w 
. e^{+\imath \Delta t } 
. e^{+\imath \left( k_j-k_{j-1}  \right) z } 
\\
&\approx&
  \left(
     -\gamma_2 + \imath \Delta + 2 \imath g' \sum_j A_j^* A_j
  \right) 
  \rho_{12}'
+ 
4 \imath f' w . 
  \mathbb{R}e
    \left[
      \sum_j A_{j}  A_{j+1}^*
      . e^{+\imath \Delta t } 
      . e^{+\imath \left( k_j-k_{j-1}  \right) z } 
    \right]
\nonumber
\\
&& ~~~~ ~~~~
+ 
4 \imath^2 f' w . 
  \mathbb{I}m
    \left[
      \sum_j A_{j}  A_{j+1}^*
      . e^{+\imath \Delta t } 
      . e^{+\imath \left( k_j-k_{j-1}  \right) z } 
    \right]
\\
\left(\textrm{split}~ \rho_{12}'\right)
~~~~
~~~~
  \partial_t u 
&=& 
 -\gamma_2 u 
 - 
  \left(
     \Delta + 2 g' \sum_j A_j^* A_j
  \right) 
  v
 -
  8 f' w . 
  \mathbb{I}m
    \left[
      \sum_j A_{j}  A_{j+1}^*
      . e^{+\imath \Delta t } 
      . e^{+\imath \left( k_j-k_{j-1}  \right) z } 
    \right]
\\
  \partial_t v 
&=& 
 -\gamma_2 v
 + 
  \left(
     \Delta + 2 g' \sum_j A_j^* A_j
  \right) 
  u
 + 
  8 f' w . 
  \mathbb{R}e
    \left[
      \sum_j A_{j}  A_{j+1}^*
      . e^{+\imath \Delta t } 
      . e^{+\imath \left( k_j-k_{j-1}  \right) z } 
    \right]
\end{eqnarray}

Where the factor $2 \imath f'$ turns into $4 \imath f'$ because the 
double summation gives two identical terms that only occur once
in the single summation.  When split into equations for $u'$ and 
$v'$, the corresponding factor becomes $8 \imath f'$.  This equation for 
$\rho_{12}'$ is equivalent to (SMN 2)\cite{Syed-MN-2000josab}, except 
I have just $k_j-k_{j-1}$
whereas they have $\Delta_j = k_j - k_{j-1} - k_0 + k_{-1}$; however
note they have the reverse sign in their carrier wave, so the only physical
difference is the $k_0 - k_{-1}$ part; also my definitions of the 
coupling differs slightly.

% ------------------------------
\subsection{Inversion ($w$)}\label{ss-multifield-w}

And now the inversion ($w$) equation (also from 
eqn.(\ref{eqn-rbpostRWA-w0-rho})) (watch for any $\omega_b$ vs $\omega_b'$ 
confusion) -- 
~
\begin{eqnarray}
\partial_t w
&=&
- \gamma_1 \left( w - w_i \right)  
+ 4 \imath f' 
   A^* A
\left[
  \rho_{12}' e^{+\imath \omega_b' t} 
-
  \rho_{12}'^* e^{-\imath \omega_b' t} 
\right]
\\
(A2) ~~
\partial_t w
&=&
- \gamma_1 \left( w - w_i \right)  
+ 4 \imath f' 
   \sum_j \sum_k A_j^*A_k 
         . e^{+\imath \left( \omega_j - \omega_k \right) t} 
         . e^{+\imath \left( -k_j + k_k  \right) z} 
\left[
  \rho_{12}' e^{+\imath \omega_b' t} 
-
  \rho_{12}'^* e^{-\imath \omega_b' t} 
\right]
\\
&=&
- \gamma_1 \left( w - w_i \right)  
+  4 \imath f' 
  \sum_j \sum_k 
\left[
  A_j^*A_k 
  \rho_{12}' 
          e^{+\imath \left( j -k +1 \right) \omega_b t - \imath \Delta t } 
-
  A_j^*A_k 
  \rho_{12}'^* 
          e^{+\imath \left( j -k -1 \right) \omega_b t + \imath \Delta t} 
\right]
         . e^{+\imath \left( k_k -k_j \right) z} 
\\(RWA) ~~
&\approx&
- \gamma_1 \left( w - w_i \right)  
+  8 \imath f' \sum_j
\left[ 
  A_j^*A_{j+1}
  \rho_{12}' 
         . e^{-\imath \Delta t }
         . e^{+\imath \left( k_{j+1} -k_j \right) z } 
-
  A_j^*A_{j-1}
  \rho_{12}'^* 
         . e^{+\imath \Delta t }
         . e^{+\imath \left( k_{j-1} -k_j \right) z } 
\right]
\\
&=&
- \gamma_1 \left( w - w_i \right)  
+  4 \imath f' \sum_j
\left[ 
  A_j^*A_{j+1}
  \left( u' + \imath v' \right)
         . e^{-\imath \Delta t }
         . e^{+\imath \left( k_{j+1} -k_j \right) z } 
\right.
\nonumber
\\
&& ~~~~ ~~~~ ~~~~ ~~~~ ~~~~ ~~~~ ~~~~ 
\left.
-
  A_{j+1}^*A_{j}
  \left( u' - \imath v' \right)
         . e^{+\imath \Delta t }
         . e^{+\imath \left( k_{j} -k_{j+1} \right) z } 
\right]
\\
&=&
- \gamma_1 \left( w - w_i \right)  
+ 
 4 \imath u' f' \sum_j
\left[ 
  A_j^*A_{j+1}
         . e^{-\imath \Delta t }
         . e^{+\imath \left( k_{j+1} -k_j \right) z } 
  - c.c.
\right]
\nonumber
\\
&& ~~~~ ~~~~ ~~~~ ~~~~ ~~~~ ~~~~ ~~~~ 
-
 4 v' f' \sum_j
\left[ 
  A_j^*A_{j+1}
         . e^{-\imath \Delta t }
         . e^{+\imath \left( k_{j+1} -k_j \right) z } 
  + c.c.
\right]
\\
&=&
- \gamma_1 \left( w - w_i \right)  
- 
 8 u' f' \sum_j
\mathbb{I}\textrm{m} \left[ 
  A_j^*A_{j+1}
         . e^{-\imath \Delta t }
         . e^{+\imath \left( k_{j+1} -k_j \right) z } 
\right]
\nonumber
\\
&& ~~~~ ~~~~ ~~~~ ~~~~ ~~~~ ~~~~ ~~~~ 
-
 8 v' f' \sum_j
\mathbb{R}\textrm{e} \left[ 
  A_j^*A_{j+1}
         . e^{-\imath \Delta t }
         . e^{+\imath \left( k_{j+1} -k_j \right) z } 
\right]
\end{eqnarray}

Note the RWA's above (for both polarization and inversion 
equations) discard modulations at frequency 
of multiples of $\omega_b$.
Quite a lot of physics has been removed by these RWA approximations, 
although it is very reasonable except in the very wideband limit.
For example, the effect of next-nearest neighbour field components
acting on the transition have been ignored, 
as have all more distant field-field interactions. 
In the next-nearest neighbour case, the dropped terms would impose a 
rapid $\omega_b$ oscillation onto the polarization $\rho_{12}$, 
which would in turn tend to impose sidebands at $\pm \omega_b$ onto
each field component.  
It is reasonable to ignore such sidebands
in the narrowband limit studied by most users of a multi-field Raman 
theory; 
but, in principle one might extend a multi-field theory to include them 
by inventing a scheme to apply the sidebands to the field component they 
are (near) resonant with.

% ------------------------------
\subsection{Fields ($A_j$)}\label{ss-multifield-A}

Note that the field evolution equation already has a carrier of 
$\exp \left[ -\imath \left( \omega_0 t - k_0 z \right) \right]$ 
factored out of it.  Thus I use $A'$ from 
$A = A' \exp \left[ -\imath \left( \omega_0 t - k_0 z \right) \right]$, 
not $A$.  Finally, I need to insert the GFEA few-cycle correction 
to the polarization term, because my ($j\ne 0$) sub-envelopes $A_j$
have an $\imath j \omega_b t$  time dependence that cannot be neglected.

From eqns.(\ref{eqn-single-Apropagate}, \ref{eqn-single-Xrho}), I get
~
\begin{eqnarray}
\partial_z A'(t)
&=&
\imath 
\left[
 1 + \frac{\imath \partial_t }{\omega_0}
\right]
\frac{\zeta \sigma \omega_0 \bar{\alpha}_{12}'}
     {2 \epsilon_0 c_0}
A'(t) X(t)
\label{eqn-single-Apropagate2-copy}
\\
\partial_z 
\sum_j A_j \exp\left[ -\imath \left( \omega'_j t - k'_j z \right) \right]
&=&
  \imath 
\left[
 1 + \frac{\imath \partial_t}{\omega_0}
\right]
  \frac{\zeta \sigma \omega_0 \bar{\alpha}_{12}'}
       {2 \epsilon_0 c_0}
  \left[
    \rho_{12}'    e^{ +\imath \omega_b' t }
   +
    \rho_{12}'^*  e^{ -\imath \omega_b' t }
  \right]
\nonumber
\\
&&
~~~~ ~~~~ ~~~~ ~~~~ ~~~~ ~~~~ ~~~~ ~~~~ ~~~~ ~~~~ \times 
  \sum_j A_j \exp\left[ -\imath \left( \omega'_j t - k'_j z \right) \right]
\\
(\textrm{match} ~ \omega_j ~ \textrm{terms})  ~~~~
  \left[
    \imath 
    k'_j 
      A_j
  +
    \partial_z A_j
  \right]
  e^{-\imath j \omega_b t }
&=&
\imath 
\frac{\zeta \sigma \bar{\alpha}_{12}'}
     {2 \epsilon_0 c_0}
  \left\{
    A_{j+1} \rho_{12}' 
      \exp\left[ +\imath (k'_{j+1} -k'_j) z - \imath \Delta t \right]
\right.
\nonumber
\\
&& ~~~~ ~~~~ ~~~~ ~~~~ ~~~~ ~~~~ 
\left.
   + 
   A_{j-1} \rho_{12}'^* 
      \exp\left( +\imath (k'_{j-1} -k'_j) z + \imath \Delta t \right)
  \right\}
\nonumber
\\
&& ~~~~ ~~~~ ~~~~ ~~~~ ~~~~ \times  ~~~~ 
\left[
 \omega_0  + \imath \partial_t
\right]
  e^{-\imath j \omega_b t }
\end{eqnarray}

Then, using
~
\begin{eqnarray}
  \left[
   \omega_0  + \imath \partial_t
  \right]
  e^{-\imath j \omega_b t }
&\longrightarrow&
   \omega_0 
  e^{-\imath j \omega_b t }
-
   \imath^2
   j \omega_b
  e^{-\imath j \omega_b t }
~~~~
\longrightarrow
~~~~
  \left[
       \omega_0 + j \omega_b
  \right]
  e^{-\imath j \omega_b t }
~~~~
\longrightarrow
~~~~
  \omega_j
  e^{-\imath j \omega_b t }
\end{eqnarray}

So
~
\begin{eqnarray}
  \left[
    \imath 
    k'_j 
      A_j
  +
    \partial_z A_j
  \right]
  e^{-\imath j \omega_b t }
&=&
\imath 
\frac{\zeta \sigma \bar{\alpha}_{12}'}
     {2 \epsilon_0 c_0}
  \left\{
    A_{j+1} \rho_{12}' 
      \exp\left[ +\imath (k'_{j+1} -k'_j) z - \imath \Delta t \right]
\right.
\nonumber
\\
&& ~~~~ ~~~~ ~~~~ ~~~~ ~~~~ ~~~~ 
\left.
   + 
   A_{j-1} \rho_{12}'^* 
      \exp\left( +\imath (k'_{j-1} -k'_j) z + \imath \Delta t \right)
  \right\}
~~~~ \times  ~~~~ 
 \omega_j
  e^{-\imath j \omega_b t }
\\
    \imath 
    k'_j 
      A_j
  +
    \partial_z A_j
&=&
\imath 
\frac{\zeta \sigma \omega_j \bar{\alpha}_{12}'}
     {4 \epsilon_0 c_0}
  u
  \left[
   A_{j+1}
      \exp\left[ +\imath (k'_{j+1} -k'_j) z - \imath \Delta t \right]
   + 
    A_{j-1}
      \exp\left[ +\imath (k'_{j-1} -k'_j) z + \imath \Delta t \right]
  \right]
\nonumber
\\
&& ~~
-
\frac{\zeta \sigma \omega_j \alpha_{12}}
     {4 \epsilon_0 c_0}
  v
  \left[
   A_{j+1}
      \exp\left( +\imath (k'_{j+1} -k'_j) z  - \imath \Delta t \right)
   -
    A_{j-1}
      \exp\left( +\imath (k'_{j-1} -k'_j) z  + \imath \Delta t \right)
  \right] 
~~~~ ~~~~
\\
    \partial_z A_j
&=&
\frac{\zeta \sigma \omega_j \alpha_{12}}
     {4 \epsilon_0 c_0}
\left\{
  -
  \left[
     v 
   -
     \imath u
  \right]
   A_{j+1}
      \exp\left( +\imath (k'_{j+1} -k'_j) z - \imath \Delta t \right)
 +
  \left[
     v 
   +
     \imath u
  \right]
    A_{j-1}
      \exp\left( +\imath (k'_{j-1} -k'_j) z + \imath \Delta t \right)
\right\}
\nonumber
\\
&& ~~~~ ~~~~ ~~~~ ~~~~ 
 - 
  \imath \left( k_j - k_0 \right) A_j
\end{eqnarray}

This is in agreement with both HPB\cite{Hickman-PB-1986pra} barring the
opposite sign on the RHS -- similar agreement occurs with
SMN\cite{Syed-MN-2000josab} once I identify $q = (v+\imath u)$.  Note that 
generally $\Delta=0$, as it just controls a frame rotation for $\rho_{12}$.

Note that we can assume, quite reasonably, that the multi-field 
envelopes $A_j$ will be better behaved than the single-field envelope $A$.
However, we have made {\em more} approximations, notably by RWA'ing away 
all the 
off-resonant cross terms driving the atomic transition so a multi-field
approach is not always better.  These off-resonant terms are
at $2\omega_0 \pm \omega_b$ -- see just prior to the starting point above of 
eqns.(\ref{eqn-rbpostRWA-w0-rho}, \ref{eqn-rbpostRWA-w0-w}).

In photon variables, the above field propagation equation is (using $R=4 \hbar \sigma$, also note $\zeta=2$ to conserve energy)
~
\begin{eqnarray}
    \partial_z A_{p,j}
&=&
  \frac{\zeta}{4}
  R
  \omega_j f_p
\left\{
  -
  \left[
     v 
   -
     \imath u
  \right]
   A_{p,j+1}
      \exp\left( +\imath (k'_{j+1} -k'_j) z - \imath \Delta t \right)
 +
  \left[
     v 
   +
     \imath u
  \right]
    A_{p,j-1}
      \exp\left( +\imath (k'_{j-1} -k'_j) z + \imath \Delta t \right)
\right\}
\nonumber
\\
&& ~~~~ ~~~~ ~~~~ ~~~~ 
 - 
  \left( k_j - k_0 \right) A_{p,j}
\end{eqnarray}

% ----------------------------------------------------------------------

%\newpage

\section{Comparisons}\label{s-comparisons}

See {\em ``Wideband pulse propagation: single-field and multi-field
approaches to Raman interactions''} 
by Kinsler and New \cite{Kinsler-N-2005pra}.

% ----------------------------------------------------------------------

%\newpage

\section{Summary}\label{s-summary}

I describe how to model a multi-frequency field such as
that seen in a wideband Raman generation experiment using a single
field envelope rather than a set of envelopes, one at each Stokes or
anti-Stokes frequency.  This requires that the field be propagated
taking into account wideband effects, as described by either the SEWA theory 
of Brabec and Krausz \cite{Brabec-K-1997prl}, or the more general 
GFEA of Kinsler and New \cite{Kinsler-N-2003pra}.

The usefulness of this single-field approach is not restricted to the 
Raman interaction described in this paper.  
It would be equally valuable for a near-degenerate optical parametric 
oscillator, or indeed any system where any two or more field 
components contain spectra that start to overlap as the pump or probe 
pulses get shorter.

It is important to note that it will usually only be more efficient 
to use a single-field simulation if pump pulses are very short, and effects
like the next-nearest neighbour field interactions, neglected in 
the multi-field theory, need to be included, or if the extra computational 
overhead is not inconvenient.
This is because in a single-field simulation, a very fine
time-resolution is necessary to model the polarization and field oscillations 
closely enough to get good numerical convergence.
However, this situation improves when the Raman transition has a 
smaller frequency compared to the pump pulse frequencies.  
One useful side effect of the fine time resolution is that it naturally
gives a wide spectral bandwidth, so that many Stokes and anti-Stokes 
lines are modeled quite naturally.  
Further, our single-field model could be invaluable in
modeling a short pulse pump-probe experiment 
where the probe frequency does not match any of the 
Stokes or anti-Stokes spectral lines generated by the pump pulse(s).  
A multi-field simulation would then need arrays for both 
the pump and probe Raman `ladders'' of Stokes/anti-Stokes lines, and 
the role of next nearest neighbour interactions (ignored in the multi-field
model) could well become more significant.

In summary, the advantages of our single-field approach are twofold.  
First, it includes more physics than the multi-field approach, 
even compared to a multi-field approach enhanced by adding GFEA corrections 
to the propagation of the field components.  
Secondly, it deals effortlessly with the complications of overlapping
spectra in the multi-field case.

% ----------------------------------------------------------------------
% ----------------------------------------------------------------------

% ----------------------------------------------------------------------
% ----------------------------------------------------------------------

\begin{thebibliography}{19}
\expandafter\ifx\csname natexlab\endcsname\relax\def\natexlab#1{#1}\fi
\expandafter\ifx\csname bibnamefont\endcsname\relax
  \def\bibnamefont#1{#1}\fi
\expandafter\ifx\csname bibfnamefont\endcsname\relax
  \def\bibfnamefont#1{#1}\fi
\expandafter\ifx\csname citenamefont\endcsname\relax
  \def\citenamefont#1{#1}\fi
\expandafter\ifx\csname url\endcsname\relax
  \def\url#1{\texttt{#1}}\fi
\expandafter\ifx\csname urlprefix\endcsname\relax\def\urlprefix{URL }\fi
\providecommand{\bibinfo}[2]{#2}
\providecommand{\eprint}[2][]{\url{#2}}

\bibitem[{\citenamefont{Kinsler and New}(2005)}]{Kinsler-N-2005pra}
\bibinfo{author}{\bibfnamefont{P.}~\bibnamefont{Kinsler}} \bibnamefont{and}
  \bibinfo{author}{\bibfnamefont{G.}~\bibnamefont{New}},
  \bibinfo{journal}{Phys. Rev. A.} \textbf{\bibinfo{volume}{72}},
  \bibinfo{pages}{033804} (\bibinfo{year}{2005}),
  \urlprefix\url{http://link.aps.org/abstract/PRA/v72/e033804}.

\bibitem[{\citenamefont{Harris and Sokolov}(1998)}]{Harris-S-1998prl}
\bibinfo{author}{\bibfnamefont{S.~E.} \bibnamefont{Harris}} \bibnamefont{and}
  \bibinfo{author}{\bibfnamefont{A.~V.} \bibnamefont{Sokolov}},
  \bibinfo{journal}{Phys. Rev. Lett.} \textbf{\bibinfo{volume}{81}},
  \bibinfo{pages}{2894} (\bibinfo{year}{1998}).

\bibitem[{\citenamefont{Sokolov et~al.}(2001)\citenamefont{Sokolov, Walker,
  Yavuz, Yin, and Harris}}]{Sokolov-WYYH-2001prl}
\bibinfo{author}{\bibfnamefont{A.~V.} \bibnamefont{Sokolov}},
  \bibinfo{author}{\bibfnamefont{D.~R.} \bibnamefont{Walker}},
  \bibinfo{author}{\bibfnamefont{D.~D.} \bibnamefont{Yavuz}},
  \bibinfo{author}{\bibfnamefont{G.~Y.} \bibnamefont{Yin}}, \bibnamefont{and}
  \bibinfo{author}{\bibfnamefont{S.~E.} \bibnamefont{Harris}},
  \bibinfo{journal}{Phys. Rev. Lett.} \textbf{\bibinfo{volume}{87}},
  \bibinfo{pages}{033402} (\bibinfo{year}{2001}).

\bibitem[{\citenamefont{Hakuta et~al.}(1997)\citenamefont{Hakuta, Suzuki,
  Katsuragawa, and Li}}]{Hakuta-SKL-2000prl}
\bibinfo{author}{\bibfnamefont{K.}~\bibnamefont{Hakuta}},
  \bibinfo{author}{\bibfnamefont{M.}~\bibnamefont{Suzuki}},
  \bibinfo{author}{\bibfnamefont{M.}~\bibnamefont{Katsuragawa}},
  \bibnamefont{and} \bibinfo{author}{\bibfnamefont{J.~Z.} \bibnamefont{Li}},
  \bibinfo{journal}{Phys. Rev. Lett.} \textbf{\bibinfo{volume}{79}},
  \bibinfo{pages}{209} (\bibinfo{year}{1997}).

\bibitem[{\citenamefont{Sali et~al.}(2004)\citenamefont{Sali, Mendham, Tisch,
  Halfmann, and Marangos}}]{Sali-MTHM-2004ol}
\bibinfo{author}{\bibfnamefont{E.}~\bibnamefont{Sali}},
  \bibinfo{author}{\bibfnamefont{K.}~\bibnamefont{Mendham}},
  \bibinfo{author}{\bibfnamefont{J.}~\bibnamefont{Tisch}},
  \bibinfo{author}{\bibfnamefont{T.}~\bibnamefont{Halfmann}}, \bibnamefont{and}
  \bibinfo{author}{\bibfnamefont{J.}~\bibnamefont{Marangos}},
  \bibinfo{journal}{Opt. Lett.} \textbf{\bibinfo{volume}{29}},
  \bibinfo{pages}{495} (\bibinfo{year}{2004}).

\bibitem[{\citenamefont{Sali et~al.}(2005)\citenamefont{Sali, Kinsler, New,
  Mendham, Halfmann, Tisch, and Marangos}}]{Sali-KNMHTM-2005pra}
\bibinfo{author}{\bibfnamefont{E.}~\bibnamefont{Sali}},
  \bibinfo{author}{\bibfnamefont{P.}~\bibnamefont{Kinsler}},
  \bibinfo{author}{\bibfnamefont{G.}~\bibnamefont{New}},
  \bibinfo{author}{\bibfnamefont{K.}~\bibnamefont{Mendham}},
  \bibinfo{author}{\bibfnamefont{T.}~\bibnamefont{Halfmann}},
  \bibinfo{author}{\bibfnamefont{J.}~\bibnamefont{Tisch}}, \bibnamefont{and}
  \bibinfo{author}{\bibfnamefont{J.}~\bibnamefont{Marangos}},
  \bibinfo{journal}{Phys. Rev. A} \textbf{\bibinfo{volume}{72}},
  \bibinfo{pages}{013813} (\bibinfo{year}{2005}),
  \urlprefix\url{http://link.aps.org/abstract/PRA/v72/e013813}.

\bibitem[{\citenamefont{Kinsler and New}(2003)}]{Kinsler-N-2003pra}
\bibinfo{author}{\bibfnamefont{P.}~\bibnamefont{Kinsler}} \bibnamefont{and}
  \bibinfo{author}{\bibfnamefont{G.}~\bibnamefont{New}},
  \bibinfo{journal}{Phys. Rev. A.} \textbf{\bibinfo{volume}{67}},
  \bibinfo{pages}{023813} (\bibinfo{year}{2003}),
  \urlprefix\url{http://link.aps.org/abstract/PRA/v67/e023813}.

\bibitem[{\citenamefont{P.Kinsler}(2002)}]{Kinsler-FCPP}
\bibinfo{author}{\bibnamefont{P.Kinsler}},
  \bibinfo{journal}{arXiv.org/physics/0212014}  (\bibinfo{year}{2002}),
  \urlprefix\url{http://arXiv.org/physics/0212014}.

\bibitem[{\citenamefont{R.M.Joseph and Taflove}(1997)}]{Joseph-T-1997itap}
\bibinfo{author}{\bibnamefont{R.M.Joseph}} \bibnamefont{and}
  \bibinfo{author}{\bibfnamefont{A.}~\bibnamefont{Taflove}},
  \bibinfo{journal}{IEEE Trans. Antennas Propag.}
  \textbf{\bibinfo{volume}{45}}, \bibinfo{pages}{364} (\bibinfo{year}{1997}).

\bibitem[{\citenamefont{Tyrrell et~al.}(2005)\citenamefont{Tyrrell, Kinsler,
  and New}}]{Tyyrell-KN-2005jmo}
\bibinfo{author}{\bibfnamefont{J.}~\bibnamefont{Tyrrell}},
  \bibinfo{author}{\bibfnamefont{P.}~\bibnamefont{Kinsler}}, \bibnamefont{and}
  \bibinfo{author}{\bibfnamefont{G.}~\bibnamefont{New}}, \bibinfo{journal}{J.
  Mod. Opt.} \textbf{\bibinfo{volume}{52}}, \bibinfo{pages}{973}
  (\bibinfo{year}{2005}),
  \urlprefix\url{http://journalsonline.tandf.co.uk/openurl.asp?genre=article&i%
d=doi:10.1080/09500340512331334086}.

\bibitem[{\citenamefont{Hickman et~al.}(1986)\citenamefont{Hickman, Paisner,
  and Bischel}}]{Hickman-PB-1986pra}
\bibinfo{author}{\bibfnamefont{A.}~\bibnamefont{Hickman}},
  \bibinfo{author}{\bibfnamefont{J.}~\bibnamefont{Paisner}}, \bibnamefont{and}
  \bibinfo{author}{\bibfnamefont{W.~K.} \bibnamefont{Bischel}},
  \bibinfo{journal}{Phys. Rev. A} \textbf{\bibinfo{volume}{33}},
  \bibinfo{pages}{1788} (\bibinfo{year}{1986}),
  \urlprefix\url{http://link.aps.org/abstract/PRA/v33/p1788}.

\bibitem[{\citenamefont{McDonald et~al.}(1998)\citenamefont{McDonald, New,
  Chan, Losev, and Luttensko}}]{McDonald-NCLL-1998jmo}
\bibinfo{author}{\bibfnamefont{G.}~\bibnamefont{McDonald}},
  \bibinfo{author}{\bibfnamefont{G.}~\bibnamefont{New}},
  \bibinfo{author}{\bibfnamefont{Y.-M.} \bibnamefont{Chan}},
  \bibinfo{author}{\bibfnamefont{L.}~\bibnamefont{Losev}}, \bibnamefont{and}
  \bibinfo{author}{\bibfnamefont{A.}~\bibnamefont{Luttensko}},
  \bibinfo{journal}{J.Mod.Opt.} \textbf{\bibinfo{volume}{45}},
  \bibinfo{pages}{1099} (\bibinfo{year}{1998}).

\bibitem[{\citenamefont{Gabor}(1946)}]{Gabor-1946jiee}
\bibinfo{author}{\bibfnamefont{D.}~\bibnamefont{Gabor}}, \bibinfo{journal}{J.
  Inst. Electr. Eng. (London)} \textbf{\bibinfo{volume}{93}},
  \bibinfo{pages}{429} (\bibinfo{year}{1946}).

\bibitem[{\citenamefont{Kien et~al.}(1999)\citenamefont{Kien, Liang,
  Katsuragawa, Ohtsuki, and Hakuta1}}]{Kien-LKOHS-1999pra}
\bibinfo{author}{\bibfnamefont{F.~L.} \bibnamefont{Kien}},
  \bibinfo{author}{\bibfnamefont{J.~Q.} \bibnamefont{Liang}},
  \bibinfo{author}{\bibfnamefont{M.}~\bibnamefont{Katsuragawa}},
  \bibinfo{author}{\bibfnamefont{K.}~\bibnamefont{Ohtsuki}}, \bibnamefont{and}
  \bibinfo{author}{\bibfnamefont{K.}~\bibnamefont{Hakuta1}},
  \bibinfo{journal}{Phys. Rev. A} \textbf{\bibinfo{volume}{60}},
  \bibinfo{pages}{1562} (\bibinfo{year}{1999}),
  \urlprefix\url{http://link.aps.org/abstract/PRA/v60/p1562}.

\bibitem[{\citenamefont{Allen and Eberly}(1975(?))}]{AllenEberly-ORTLA}
\bibinfo{author}{\bibfnamefont{L.}~\bibnamefont{Allen}} \bibnamefont{and}
  \bibinfo{author}{\bibfnamefont{J.}~\bibnamefont{Eberly}},
  \emph{\bibinfo{title}{Optical Resonance and Two--Level Atoms}}
  (\bibinfo{publisher}{Dover Publications, Inc.}, \bibinfo{year}{1975(?)}).

\bibitem[{\citenamefont{Shverdin et~al.}(2004)\citenamefont{Shverdin, Walker,
  Yavuz, Yin, and Harris}}]{Shverdin-WYYH-2004pre}
\bibinfo{author}{\bibfnamefont{M.}~\bibnamefont{Shverdin}},
  \bibinfo{author}{\bibfnamefont{D.}~\bibnamefont{Walker}},
  \bibinfo{author}{\bibfnamefont{D.}~\bibnamefont{Yavuz}},
  \bibinfo{author}{\bibfnamefont{G.}~\bibnamefont{Yin}}, \bibnamefont{and}
  \bibinfo{author}{\bibfnamefont{S.}~\bibnamefont{Harris}},
  \bibinfo{journal}{Phys. Rev. Lett.} \textbf{\bibinfo{volume}{93}},
  \bibinfo{pages}{033904} (\bibinfo{year}{2004}).

\bibitem[{\citenamefont{Seyed et~al.}(2000)\citenamefont{Seyed, McDonald, and
  New}}]{Syed-MN-2000josab}
\bibinfo{author}{\bibfnamefont{K.}~\bibnamefont{Seyed}},
  \bibinfo{author}{\bibfnamefont{G.}~\bibnamefont{McDonald}}, \bibnamefont{and}
  \bibinfo{author}{\bibfnamefont{G.}~\bibnamefont{New}}, \bibinfo{journal}{JOSA
  B} \textbf{\bibinfo{volume}{17}}, \bibinfo{pages}{1366}
  (\bibinfo{year}{2000}).

\bibitem[{\citenamefont{Brabec and Krausz}(1997)}]{Brabec-K-1997prl}
\bibinfo{author}{\bibfnamefont{T.}~\bibnamefont{Brabec}} \bibnamefont{and}
  \bibinfo{author}{\bibfnamefont{F.}~\bibnamefont{Krausz}},
  \bibinfo{journal}{Phys. Rev. Lett.} \textbf{\bibinfo{volume}{78}},
  \bibinfo{pages}{3282} (\bibinfo{year}{1997}).

\bibitem[{\citenamefont{McDonald et~al.}(1994)\citenamefont{McDonald, New,
  Losev, Luttensko, and Shaw}}]{McDonald-NLLS-1994ol}
\bibinfo{author}{\bibfnamefont{G.}~\bibnamefont{McDonald}},
  \bibinfo{author}{\bibfnamefont{G.}~\bibnamefont{New}},
  \bibinfo{author}{\bibfnamefont{L.}~\bibnamefont{Losev}},
  \bibinfo{author}{\bibfnamefont{A.}~\bibnamefont{Luttensko}},
  \bibnamefont{and} \bibinfo{author}{\bibfnamefont{M.}~\bibnamefont{Shaw}},
  \bibinfo{journal}{Opt. Lett.} \textbf{\bibinfo{volume}{19}},
  \bibinfo{pages}{1400} (\bibinfo{year}{1994}).

\end{thebibliography}
\end{document}